\documentclass[final,1p,times]{elsarticle}

\usepackage{graphicx}

\usepackage{amssymb}

\usepackage{float}

\usepackage[colorlinks,hyperindex]{hyperref}

\journal{Astroparticle Physics}

\begin{document}

\begin{frontmatter}



\title{Composite macro-bolometers for the rejection of surface radioactive background in rare-event experiments}

\author[LAL,INS]{Luca~Foggetta} 
\author[INS]{Andrea~Giuliani\corref{cor1}}
\cortext[cor1]{Corresponding author. Universit\`a dell'Insubria, Dipartimento di Fisica e Matematica, Via Valleggio 11, I-22100 Como, Italy. Tel. +390312386217  }
\ead{andrea.giuliani@mib.infn.it}  
\author[CEA,INS]{Claudia~Nones} 
\author[LNL,INS]{Marisa~Pedretti}  
\author[INS]{Chiara~Salvioni} 
\author[LNL,INS]{Samuele~Sangiorgio} 
\address[INS]{Universit\`a dell'Insubria,  Dipartimento di Fisica e Matematica, Via Valleggio 11, 22100 Como, Italy}
\address[LAL]{Laboratoire de l'Acc\'el\'erateur Lin\'eaire, CNRS-Orsay, 91898 Orsay cedex, France}
\address[CEA]{Institut de Recherche sur les lois Fondamentales de l'Univers, CEA-Saclay, 91191 Gif sur Yvette Cedex, France}
\address[LNL]{Lawrence Livermore National Laboratory, 7000 East Avenue, Livermore, CA 94550-9234, USA}

\begin{abstract}

Experiments searching for rare events, such as neutrinoless double beta decay and interactions of dark matter candidates, require extremely low levels of background. When these experiments are performed using macro-bolometers, radioactive contamination near the surfaces is of particular concern. For a bolometric neutrinoless double beta decay experiment, it can cause counts in the spectral region where the signal is expected, while for a dark matter experiment which exploits ionization signals for particle identification, it originates an incomplete charge collection simulating a nuclear recoil.  
In order to control the effects of surface contamination, we developed a novel technique that uses composite macro-bolometers to identify energy depositions that occur close to the surfaces of materials immediately surrounding the detector.  The composite macro-bolometer proposed and studied here consists of a main energy absorber that is thermally coupled to and entirely surrounded by thin absorbers that act as active shields. Surface energy depositions can be rejected by the analysis of simultaneous signals in the main absorber and the shields.   
In this paper, we describe a full thermal model and experimental results for three prototype detectors. The detectors consist of Ge, Si, or TeO$_2$ thin absorbers as active shields, all with TeO$_2$ crystals as main absorbers.
In all cases, the surface event rejection capability is clearly demonstrated. In addition, simulations and preliminary results show that it is possible to detect energy depositions that occurred on the shields without separate readout channels for them. The energy depositions in the shields are distinguished from those in the main absorber through pulse shape discrimination. This simplification makes this technique a viable method for the rejection of surface energy depositions in next-generation bolometric double beta decay searches, such as possible extensions or upgrades of the CUORE experiment.

\end{abstract}


\end{frontmatter}


\section{Introduction and motivation}
\label{sec:introduction}

The searches for the neutrino mass and dark matter are at present two of the most relevant and exciting fields in cosmology and particle physics. Experiments that search for neutrinoless double beta decay ($0\nu\beta\beta$)~\cite{rev_DBD} or the nuclear recoils induced by a WIMP \cite{rev_DM} require the detection of very rare events. Although the signals are expected in two very different energy regions, $\sim$~MeV for the former and $< 100$~keV for the latter, the experimental approaches to these searches often share common technological challenges. In both cases, bolometers are used as detectors for many sensitive experiments, such as EDELWEISS \cite{EDELWEISS}, CRESST \cite{CRESST} and CDMS \cite{CDMS} for dark matter searches or Cuoricino \cite{Cuoricino} and CUORE \cite{CUORE} for neutrinoless double beta decay.

Bolometers are phonon-mediated particle detectors operated at low temperatures~\cite{A1}. These devices are capable of obtaining both higher energy resolutions and lower energy thresholds than conventional detectors. In addition, they can be fabricated from a wide variety of materials, allowing flexibility for experiments that require the detectors to contain particular atomic or nuclear species. If other excitations (such as ionization charge carriers or scintillation phonons) are exploited in addition to phonons, bolometers can discriminate nuclear recoils from electron recoils, or $\alpha$~particles from $\beta$~particles and $\gamma$~rays.

Bolometer-based $0\nu\beta\beta$ and dark matter experiments require extremely low levels of radioactive background, since they search for very rare events. For different reasons in the two cases, surface radioactive contamination is of particular concern. In $0\nu\beta\beta$ searches, $\alpha$'s and $\beta$'s of superficial origin can lose part of their energy in a few microns and generate a continuum in the spectral region where the $0\nu\beta\beta$ signal is expected. In dark matter searches, electron recoil at the detector surface (from any type of ionizing interactions) can result in incomplete charge collection, mimicking a nuclear recoil. Therefore, a bolometer capable of tagging surface events would be a powerful tool to improve the signal-to-background ratio in both types of experiments.

In monolithic bolometers it is possible to obtain some spatial information by detecting phonons before they thermalize. This is used in the dark matter experiments CDMS and EDELWEISS~\cite{A5,NBSI}. Recently, it was shown that this technique can be extended to bolometric detectors for $0\nu\beta\beta$ as well, though incomplete thermalization leads to degradation in energy resolution~\cite{NONES-LTD12,NONES-PHD}. For dobule beta decay experiments, it is important to have good energy resolution to improve the signal-to-background ratio. In this case, it is preferable to operate the bolometers in calorimetric mode, where the phonon thermalization from an event is almost complete and the deposited energy is fully converted to heat and measured as a temperature rise. However, this mode loses all spatial information about the event, which makes rejection of surface background events more difficult. 

\begin{figure}[tbp]
\centering
\includegraphics[width=0.4\textwidth]{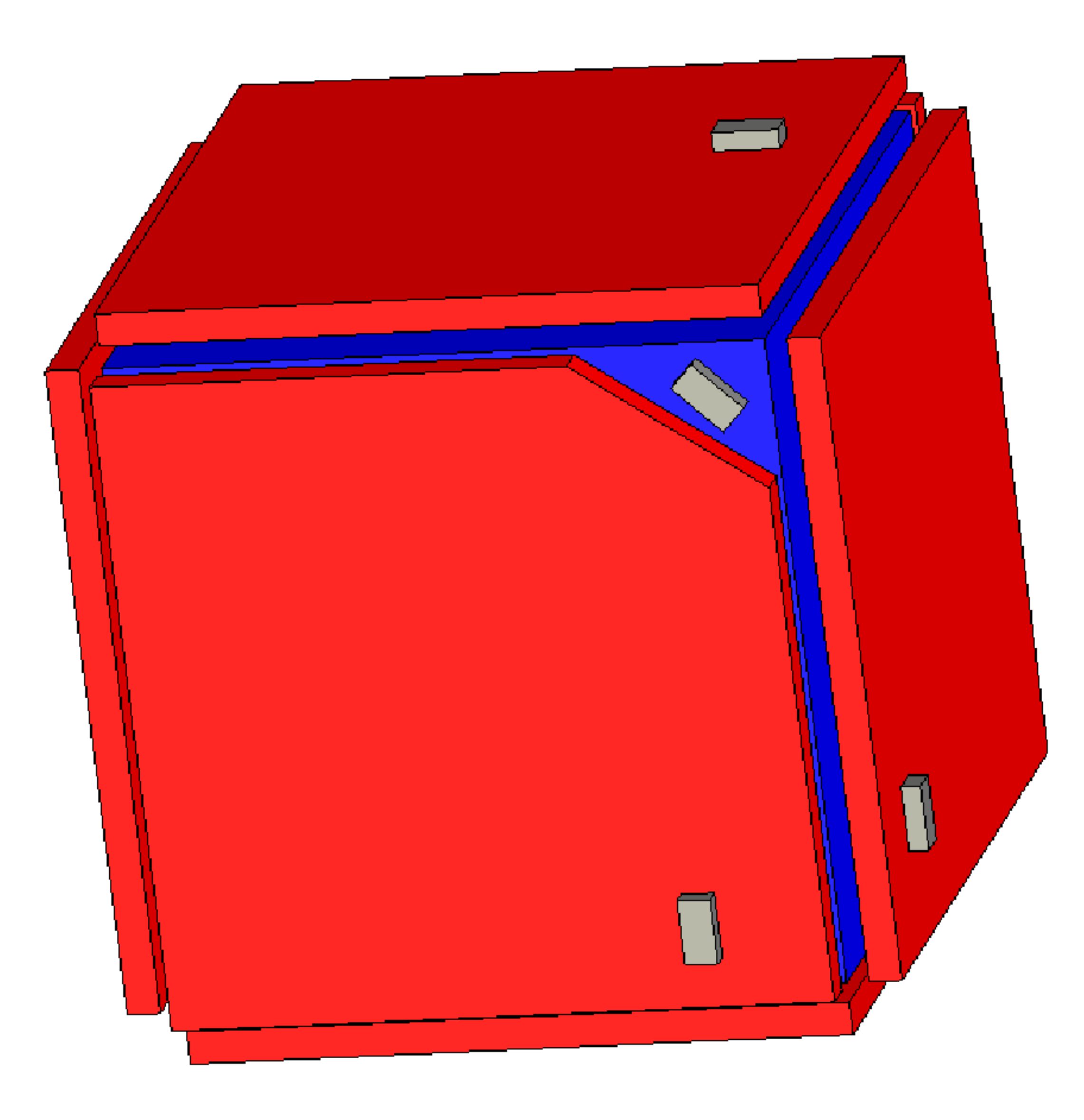}
\caption{Schematic 3D view of a surface sensitive bolometer. The main absorber is shown in dark gray (blue) while the thin auxiliary bolometers are in light gray (red) and are thermally connected to the main absorber. The small elements on each absorber surface are the temperature sensors.}
\label{ssb-3D}
\end{figure}

The aim of this study is to demonstrate that the lack of spatial resolution intrinsic to low temperature calorimeters may be partially overcome by utilizing a composite bolometer consisting of a main energy absorber that is thermally coupled to and surrounded by thin absorbers working as auxiliary bolometers and active shields (Fig. \ref{ssb-3D}). We call this composite bolometer a ``surface sensitive bolometer'' or SSB. 

The motivation of this work originates in physics experiments being carried out at the underground Gran Sasso National Laboratory (LNGS) in Italy searching for neutrinoless double beta decay of $^{130}$Te and named Cuoricino~\cite{Cuoricino} and CUORE~\cite{CUORE}. Cuoricino ended in June 2008 and was constituted by 62 TeO$_2$ crystals of two sizes, $\sim330$~g and $\sim790$~g, with a total of 40.7\,kg of TeO$_2$. CUORE is a next generation experiment currently under construction, and will have 988 crystals for a total of 741\,kg of TeO$_2$. Cuoricino was operated at 10~mK, and CUORE is designed to operate at similar temperatures.  These experiments look for a peak in the energy spectrum at 2527~keV \cite{pen-1,pen-2}. The observation of $0\nu\beta\beta$ peak  would imply that neutrinos are self-conjugate particles and enables a sensitive measurement of their mass scale. These experiments are looking for events of less than 10$^{-2}$~counts/y/mole and it is critical to control the background generated by residual radioactive impurities both inside and near the detector. A background of at least 0.01 counts/keV/kg/y near the energy region of interest is required for CUORE to reach its desired sensitivity. 

Simulations based on the background levels observed in Cuoricino and measurements of radioactive contaminants with germanium detector gamma counting and other techniques show that $\sim$ 60\% of the contribution to the radioactive background at the energy of interest can be identified as partial energy deposition of $\alpha$ and $\beta$-particles emitted from either the surface of the detectors or the materials that surround them \cite{Cuoricino}. These particles release only a fraction of their energy into the bolometer, and produce a roughly continuous spectrum which extends from the full energy of the decay to zero. With respect to Cuoricino, a substantial improvement was obtained in R\&D tests performed in view of CUORE~\cite{Gorla}. In particular, the contribution coming from TeO$_2$ crystal surfaces was reduced thanks to a new polishing procedure~\cite{DAF} and is now compatible with the aforementioned target of 0.01 counts/keV/kg/y. On the contrary, the background coming from the inert material surfaces around the detector, in particular the Cu holding structure, has not been reduced yet at the desired level. The work described here aims at decreasing this background by developing detectors that are capable of identifying these surface events~\cite{A12,PHDSAM}.

The present paper is organized as follows. In Section~\ref{sec:structure} the basic principles and structure of the proposed detectors are described. Afterwards (Section~\ref{sec:model}), we present a thermal model of the SSB that we use to simulate its performance. The rest of the paper describes our prototype detectors equipped with SSBs and their experimental characterization. After a description of the setup in Section~\ref{sec:setup}, the experimental data are shown and discussed in Section~\ref{sec:results} with a presentation of two approaches for the identification of surface events and a preliminary quantification of the rejection power. Section~\ref{sec:dt} is dedicated to a further approach that promises to tag surface events with a passive use of the shields. Preliminary encouraging results and simulations are presented. Conclusions and prospects are discussed in Section~\ref{sec:conclusion}. 

\section{Basic structure and operation principle of Surface Sensitive Bolometers}
\label{sec:structure}

A bolometer is a solid-state particle detector composed of two elements: an energy absorber, where the energy of the particles is deposited and converted into phonons, and a phonon sensor attached to the absorber. The particle energy is initially stored in the form of optical and Debye-energy  phonons. If the absorber is kept at very low temperatures (below 100~mK), the phonons created by the particle interaction have much higher energies than the phonon thermal bath and, in bolometers, represent the elementary excess excitations which mediate the particle detection. For macro-bolometers with dimensions of a few centimeters as those considered here, the phonon thermalization time is typically of the order of few microseconds. If either the phonon response or the phonon transmission to the sensor is slower, then the deposited energy has enough time to be converted to heat and the phonon sensor interprets a particle energy deposition as a temperature rise, i.e. the phonon sensor effectively acts as a thermometer. In the present work, phonon sensors are glued with thick (up to 50~$\mu$m) epoxy layers at the energy absorber, introducing an intrinsically slow ($\sim$~1-10~ms scale) transmission interface. The bolometer is then operated in a calorimetric mode which provides the optimal energy resolution.

Resistive elements with a strong temperature dependence are often used as thermometers. There are two types of thermometers typically used for these experiments: semiconductor thermistors, like the ones used in this study, or transition edge sensors (TES), which are superconducting films kept at their critical temperature. To minimize heat capacity, bolometers are typically operated at cryogenic temperatures below 100 mK (in some cases, less than 15 mK). Dielectric diamagnetic crystals have low heat capacities and are commonly used as energy absorbers. With such devices, energy resolutions as low as 5~eV have been achieved for X-rays~\cite{A2} and 5~keV for $\alpha$-particles~\cite{A3}; $\gamma$-rays can be detected with resolutions comparable to those obtained with the best germanium diodes~\cite{A4}.

In calorimetric operation, no difference is expected for the detector response to interactions of different ionizing particles, in particular electrons and $\alpha$-particles. However, if a relevant fraction of athermal phonons contributes to the signal formation, in principle differences between electrons and $\alpha$-particles can arise, since the different track structure and ionization density could influence the heating process. This aspect was studied experimentally by comparing the response of TeO$_{2}$ bolometers to $\gamma$-rays and $\alpha$-particles~\cite{ALE97}. It was found that the amplitudes of the signals provided by a glued semiconductor thermistor are the same within 2\% for the same energy deposition of $\gamma$-rays and $\alpha$-particles. No appreciable difference in the signal shapes was observed. On the basis of this result, the detector responses to electrons and $\alpha$-particles will be considered to be the same in the following discussion.

The basic idea of the SSB is to surround the main absorber with thin auxiliary absorbers which function as active shields and are operated as bolometers. In this study the main absorber and the shields are thermally connected, and a near 4$\pi$ coverage from external charged particles can be achieved. The SSB is designed to be able to distinguish among three types of events: those induced by particles originating outside of the entire detector and that are stopped in the shields depositing only in them their residual energy, those induced by particles originating on the surface of the main or auxiliary absorber with energy deposited in both, and those with energy deposited in the bulk of the main absorber. We refer to the first case as ``surface events'', the second one as ``mixed events'' and the third one as ``bulk events''.
 
The main absorber and the active shields may be made of the same or different materials depending on purity requirements and appropriate thermal properties. Because the main interest of this study is to reduce the background for CUORE, the main absorber proposed here and studied in the model is made of TeO$_2$ and is cubic or rectangular in shape.  Each of its six surfaces will be covered with an active shield, the area of which is similar to that of the corresponding face. In order to not significantly alter the detector geometry and space occupancy, the shields have a thickness of a fraction of a millimeter, about two orders of magnitude less than the main absorber dimension. Active shields of Si, Ge, and TeO$_2$, are considered. The choice of these materials, as discussed more extensively in Section~\ref{sec:material}, was mainly driven by reasons of radiopurity (Ge), availability and cost (Si), and thermal compatibility (TeO$_2$).In the current design, the shields are thermally and mechanically coupled to the main absorber by epoxy beads of thickness less than $\sim$~50~$\mu$m. Other coupling methods between shields and main absorber can be devised, aiming at a better control of the thermal conductance between these elements. One of them is implemented in this work and described in Section~\ref{sec:detectors}.

The thermal pulses from energy deposition in the absorbers are read out by thermistors thermally coupled to the absorbers by epoxy beads. The bolometers described in this work use semiconductor thermistors operated in the variable range hopping (VRH) conduction mode \cite{A10}. These thermistors consist of neutron transmutation doped (NTD) Ge single crystals, with a mass of the order of $\sim$~10~mg. The doping technique~\cite{A13} allows to achieve uniform doping in the whole thermistor volume and to accurately tune the net dopant concentration, corresponding to the desired temperature dependance of the resistivity at low temperatures. 

The main absorber and each of the six shields have their own thermistor, and each absorber-thermistor pair can be regarded as an individual detector. The bias on the thermistors results in a static heat flow and therefore a temperature difference between the thermistor and the absorber appears. These differences are generally much smaller than the absolute temperatures of the detector components. The main absorbers are thermally linked to a copper frame acting as a heat sink through a set of PTFE (Polytetrafluoroethylene) stand-offs, and the copper frame is thermally coupled to the mixing chamber, the coldest point of a dilution refrigerator~\cite{LOU}.

The main absorber, epoxy beads, and thermistors are the same as those used in Cuoricino, and their thermal properties measured for Cuoricino have been used as input for the simulations described in Section~\ref{sec:model}. 

Charged particles from materials outside the SSB, such as $\alpha$-particles, will be stopped and tagged by one of the active shields (surface events). They release part to most of their energy in the shield, but raise the temperature of all the detector elements as they are, in fact, all thermally connected. Because of the small heat capacity of the shield due to its small mass, the signal read by its thermistor will have a higher amplitude and faster rise time than the signal read by the thermistor attached to the main absorber. If, on the other hand, an energy deposition occurs inside the main absorber (bulk event), all of the thermistors will read pulses with comparable amplitudes and rise times. The rise time of the temperature pulse in the active shield thermistor is much slower when energy is released in the bulk of the main absorber as opposed to when it is released in the shield. It is therefore possible to separate bulk and surface events by comparing amplitude and shape of pulses among the different thermistors. No degradation in energy resolution of the main absorber is expected since the device is still operating in the mode where phonon thermalization is nearly complete. 

A specific discussion is required for $\alpha$-particles that originate from surface contamination of the main absorber or of the shield itself. As already pointed out, one of the largest backgrounds for neutrinoless double beta decay experiments like Cuoricino comes from surface contaminants that emit $\alpha$-particles. The $\alpha$-particles can mimick a $0\nu\beta\beta$~event if they deposit a part of their energy elsewhere and about 2530~keV in the main absorber, where the double beta decay peak is expected. By using the active shields discussed here, $\alpha$'s originating from outside the detectors or from the external surfaces of the shield can be distinguished from the bulk events in the main absorbers. As for $\alpha$'s escaping from the internal surfaces of the shield or from the energy absorber surface, and delivering in the main crystal about 2.5 MeV (close to the double beta decay signal), they will surely deliver at least 1.5~MeV in the shield (their remaining energy), since natural $\alpha$'s have energies higher than $\sim$4~MeV. These mixed events can be easily recognized by the large fast pulses associated that are characteristic of the energy deposited in the shields. We point out however that, as discussed in Section~\ref{sec:introduction}, the background originating from the TeO$_2$ surface is now under control thanks to appropriate cleaning methods.

\section{Thermal model and expected performances of Surface Sensitive Bolometers}
\label{sec:model}

\begin{figure}[tbp]
\centering
\includegraphics[width=0.75\textwidth]{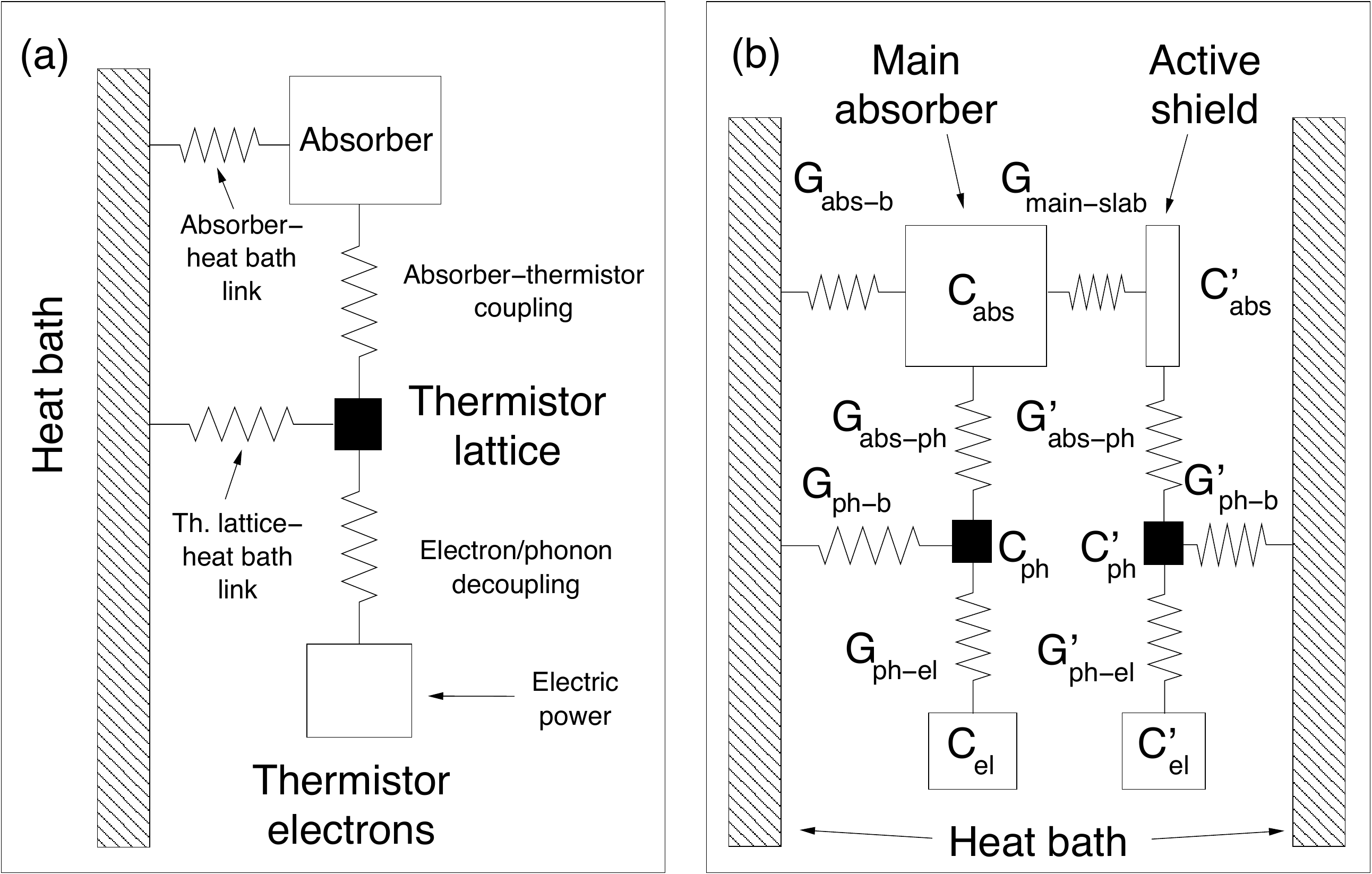}
\caption{(a) Three-node thermal model for a bolometer, consisting of the main absorber node and the two
thermistor nodes. The thermistor is separated into electron and phonon systems,
connected by the electron-phonon thermal conductance. (b) Thermal network used for the case of
a single-shield SSB. The presence of the shield adds three nodes, corresponding to those of the shield-thermistor system. In our experiment, epoxy beads are normally used to obtain heat conduction
between the main absorber and the shield ($G_{main-shield}$) and between the thermistors and the absorbers ($G_{abs-ph}$, $G'_{abs-ph}$). In addition, PTFE holders link the main absorber to the heat bath ($G_{abs-b}$) and gold wires provide electric read-out and thermal connection between thermistor lattice and heat bath ($G_{ph-b}$, $G'_{ph-b}$). $C$ and $C'$ are the heat capacities of each element of the main and of the shield bolometer respectively.}
\label{01}
\end{figure}

For a thermal detector without the active shields, the thermal network consists of three thermal nodes: the main energy absorber (here a TeO$_2$ crystal), the thermistor lattice, and the thermistor electrons (Fig.~\ref{01}a). The thermistor is split into two elements because it has been observed that in these devices the lattice phonon system and the conduction electron system behave as separate thermal stages that reach different temperatures and are connected by a finite thermal conductance, which is internal to the thermistor itself. Such decoupling is due to a non-ohmic effect explained by the so-called ``hot electron model'' \cite{A9}. The electron-phonon thermal conductance of thermistors of the same type as those used in this work was measured. The methods and the results of these measurements are reported elsewhere~\cite{A11, PED}. It has also been shown experimentally that there is no direct connection between the thermistor electron system and the heat sink \cite{A11}. 

In Fig.~\ref{01}a, the presence of heat conductors in the system is shown. As described in section \ref{sec:structure}, the heat conductors between thermistors and absorbers are epoxy beads, while PTFE mechanical supports for the main absorber act as heat conductors between the main absorber and the heat sink. The thermistor read-out wires connect the thermistor lattice system to the heat sink.

The thermal model for the SSB is an extension of the three-node model and requires three additional nodes for each active shield used in the system~\cite{PHDMAR}. For simplicity, the network adopted here includes the absorber and thermistor for the main detector and one active shield (Fig.~\ref{01}b). In general, the heat conductor between the main absorber and the active shield is again a set of epoxy beads, although other solutions can be devised (see Section~\ref{sec:detectors}).

In a static condition with external power $\pi_i$ injected into node $i$ and power $P_{ij}$ flowing between nodes $i$ and $j$, the power balance is given by:
\begin{equation}
\label{eq:01}
\pi_i - \sum_{j=0}^{6} P_{ij} = 0 \qquad i = 1, 2, \dots, 6
\end{equation}
where
\begin{equation}
\label{eq:02}
P_{ij} = \int^{T_i}_{T_j} G_{ij}(T) dT 
\end{equation}
Here, $T_i$ is the equilibrium temperature of node $i$, and $G_{ij}$ is the temperature-dependent thermal conductance between nodes $i$ and $j$, with $G_{ij} = G_{ji}$ and $G_{ii} = 0$. $G_{ij} = 0$ whenever there is no physical direct connection between nodes $i$ and $j$. The node $j=0$ corresponds to the heat bath. $\pi_i$ also includes the parasitic power dissipated in the absorber (mainly due to mechanical frictions between the main absorber and the PTFE supports) and in the electron system (due probably to parasitic currents induced by EM interferences). The thermistors are biased with constant current, and the detector as a whole reaches thermal equilibrium with the six thermal stages at different temperatures.  The electric power dissipated in the thermistors from the bias current is also included in the $\pi_i$ for the electron system of the thermistor. For well fabricated and shielded bolometers, the dominant power dissipation is ascribable to the bias current.

The dynamic behavior of the detector must be determined to evaluate the detector response to energy deposited by particles interacting either in the main absorber or in the active shield. The corresponding differential equation for node $i$ is
\begin{equation}
\label{eq:03}
C_i(T_i)\cdot\dot T_i = \pi_i - \sum_{j=0}^{6} P_{ij} \qquad i = 1, 2, \dots, 6
\end{equation}
where $C_i$ is the heat capacity of node $i$. An explicit dependence of the term $\pi_i$ on the node temperature $T_i(t)$ has to be introduced for thermistor nodes connected to a bias circuit to account for a characteristic mechanism known as \emph{electrothermal feedback}. This additional effect is caused by power dissipation by the bias current which raises the thermistor temperature and acts back on its resistance until an equilibrium is reached. This mechanism, responsible for the deviation of thermistor voltage-current curve from linearity, is included in the calculations.  

The time evolution of a heat pulse is computed by solving a set of differential equations with the initial conditions provided by the solution of the static problem and by the temperature rise due to the energy deposited by the absorbed particle. Temperature pulses on the thermistor-electron systems of both main absorber and active shield are then converted into voltage pulses using the resistance vs. temperature behavior of the thermistors parameterized by
\begin{equation}
\label{eq:04}
R(T)=R_0~exp[(T_0/T)^{0.5}]
\end{equation}
as predicted by VRH with Coulomb gap~\cite{A10}. The conversion is possible because of the constant current $I$ flowing in the thermistor. In fact, a resistance variation $\Delta R$ due to a temperature change $\Delta T$ of the electron system results in a voltage variation $\Delta V$=$I \cdot \Delta R$=$I \cdot (\partial R/ \partial T) \cdot \Delta T$. In the experiements, typical pulse amplitudes are of the order of several tens of microvolts per energy deposition of 1 MeV. These voltage pulses are amplified to a range of a few volts by a low noise voltage amplifier.

Numerical codes were developed for calculations of both static and dynamic behaviors. Values for thermal conductances and heat capacities were obtained experimentally from previous measurements with detectors similar to those used in Cuoricino~\cite{A11}. A set of simulations was performed to determine the dependance of the discrimination efficiency on several experimental parameters. The simulations are based on the same detector scheme: one TeO$_2$ main absorber, and one shielding element.  A pair of thermistors with the same $R_0$ and $T_0$ were used for the main absorber and its shield.  

We have evaluated the behavior of a detector with the configuration adopted for Cuoricino and give the relevant thermal properties in Table~\ref{tab:02}. We have simulated a Si active shield attached to the main TeO$_2$ crystal with a surface area of 5$\times$5~cm$^2$ and a thickness of 300~$\mu$m. The Cuoricino TeO$_2$ crystal is a cube with 5~cm side length. The same size is foreseen for the CUORE detectors.

\begin{table*}[tbp]
\caption{Detector parameters $C$, $G$ for Eqs.(\ref{eq:01}-\ref{eq:03}) used in the simulations. The simulated SSB, represented by the thermal network in Fig.~\ref{01}b, has a TeO$_2$ cubic main absorber with 5~cm side length and a single silicon active shield with 5$\times$5~cm$^2$ surface and 300~$\mu$m thickness. $T$ is given in Kelvin. The heat sink is assumed to be at the base temperature $T_b$=9~mK. The $R_{0}$, $T_{0}$ parameters are those used in Eq.(\ref{eq:04}).}
\label{tab:02} 
\begin{center}
\begin{tabular}{c c c c c c}
\hline 
\multicolumn{1}{|c}{$G_{abs-b}$} & $G_{main-shield}$ & $G_{abs-ph}$, $G$'$_{abs-ph}$ & $G_{ph-b}$, $G$'$_{ph-b}$ & $G_{ph-el}$, $G$'$_{ph-el}$ & \multicolumn{1}{c|}{} \\
\multicolumn{1}{|c}{[W/K]} & [W/K] & [W/K] & [W/K] & [W/K] & \multicolumn{1}{c|}{} \\
\hline
\multicolumn{1}{|c}{4$\times$10$^{-5}$$\cdot$T$^2$} & 1.3$\times$10$^{-3}$$\cdot$T$^3$ & 2.34$\times$10$^{-3}$$\cdot$T$^3$ &
9.6$\times$10$^{-5}$$\cdot$T$^{2.4}$ & 7.02$\times$10$^{-1}$$\cdot$T$^{4.37}$ & \multicolumn{1}{c|}{} \\
\hline
& & & & \\
\hline
\multicolumn{1}{|c}{$C_{el}$, $C$'$_{el}$} & $C_{abs}$ & $C$'$_{abs}$ & $C_{ph}$, $C$'$_{ph}$ & \multicolumn{1}{||c}{$R_0$} & \multicolumn{1}{c|}{$T_0$} \\  
\multicolumn{1}{|c}{[J/K]} & [J/K] & [J/K] & [J/K] & \multicolumn{1}{||c}{[$\Omega$]} & \multicolumn{1}{c|}{[K]} \\
\hline
\multicolumn{1}{|c}{9.9$\times$10$^{-9}$$\cdot$T} & 2.25$\times$10$^{-3}$$\cdot$T$^3$ & 4.6$\times$10$^{-7}$$\cdot$T$^{3}$  &2.7$\times$10$^{-8}$$\cdot$T$^3$ & \multicolumn{1}{||c}{1.5} & \multicolumn{1}{c|}{3} \\				\hline
\end{tabular}
\end{center}
\end{table*}

\begin{figure}[tbp]
\centering
\includegraphics[width=1.0\textwidth]{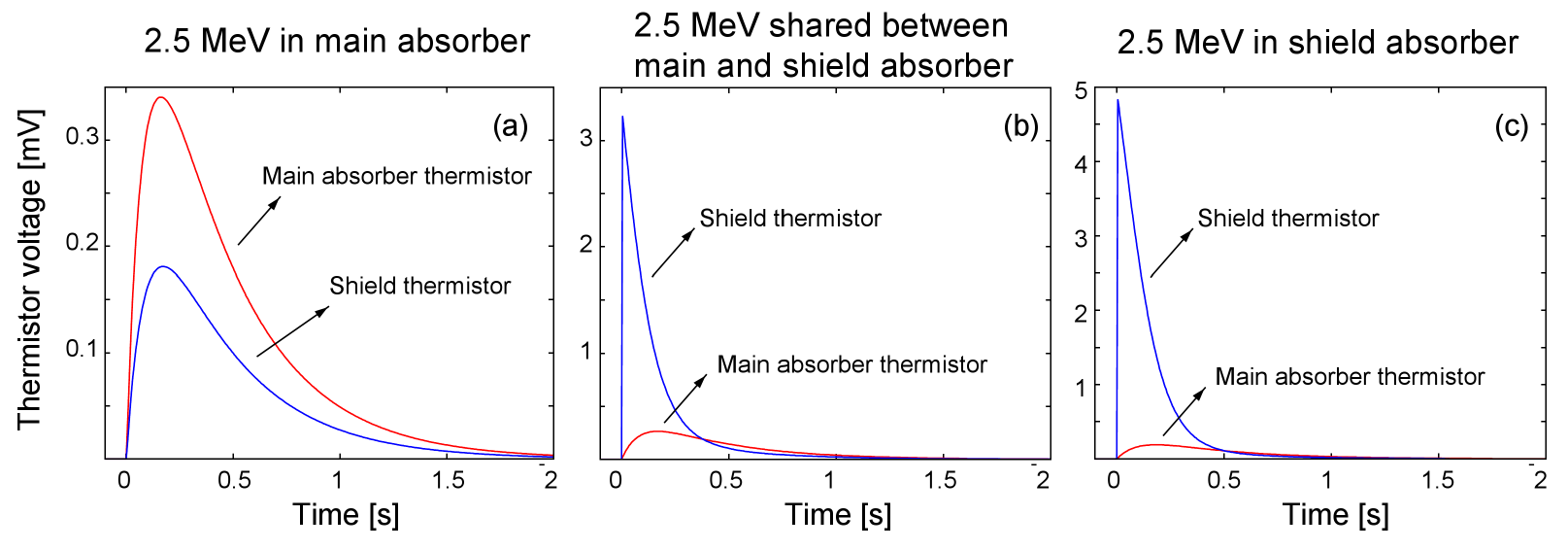}
\caption{(a) Pair of simulated pulses generated by an energy deposition of 2.5~MeV in the main absorber of the SSB. The pulses, which are read by the thermistor of the main absorber and by the thermistor of the active shield, present comparable amplitudes and shapes. The temperature increases corresponding to the voltage pulses are $5 \cdot 10^{-2}$~mK for the main absorber and $3 \cdot 10^{-2}$~mK for the shield. (b) Pair of simulated pulses generated
by a simultaneous energy deposition of 1.25~MeV in the shield of the SSB and of 1.25 MeV in the main absorber. (c) Pair of simulated pulses generated
by an energy deposition of 2.5~MeV in the shield of the SSB. Compared with the main absorber, the signal seen by the shield
thermistor is higher and faster due to the shield lower heat capacity. The temperature rises corresponding to the voltage pulses are $3 \cdot 10^{-2}$~mK for the main absorber and $1.9$~mK for the shield.}
\label{02}
\end{figure}

Fig.~\ref{02} shows three pairs of simulated pulses. They are read by thermistors on both the main absorber and the active shield simultaneously, following a specified energy deposition in the detector. Fig.~\ref{02}(a) shows simultaneous pulses for a 2.5~MeV energy deposition in the TeO$_2$ crystal (bulk event), Fig.~\ref{02}(b) shows pulses from  an energy deposition of 2.5~MeV shared in equal parts between shield and absorber (mixed event), and Fig.~\ref{02}(c) shows pulses from a 2.5~MeV energy deposition in the shield only (surface event).  The first pulse corresponds to the signal, while the second and third ones to background. The simulated pulses from the main absorber are consistent with those observed in Cuoricino both in amplitude and time structure: the amplitude is $\sim$~140~$\mu$V/MeV, the rise time (10\%--90\%) is $\tau_r$~$\sim$~50~ms, and the decay time (90\%--30\%) is $\tau_d$~$\sim$~500~ms~\cite{PHDMAR}. 

When the energy is deposited in the main absorber, the pulse decay time $\tau_d$ is $\sim C_{abs} / G_{abs-b}$, and is of the order of $\sim 500$~ms at 10~mK. (Here and in the following the symbols for the thermal parameters are the same used in Fig.~\ref{01}.) The corresponding rise time $\tau_r$ is given by $C_{parallel}/G_{ph-el}$, where $C_{parallel} = C_{abs} \cdot C_{el} /( C_{abs} + C_{el} )$. This time constant can be approximated by $C_{el} / G_{ph-el} \sim$~80~ms. If the energy is released in the active shield, $C_{abs}$ is replaced by the heat capacity of the shield $C$'$_{abs}$, two orders of magnitude lower than that of the main absorber. The ratio of the main-absorber $\tau_r$ to shield $\tau_r$ is $(C_{abs}/C$'$_{abs}) \cdot ( C$'$_{abs} + C_{el} )/( C_{abs} + C_{el} )$, which can be approximated, considering the values of the heat capacities in our model, with $C_{el} / C$'$_{abs}$: this is of the order of 10$^3$ at 10 - 15 ~mK. These time constants, determined with simple approximations, are in reasonable agreement with the detailed simulation results shown in Fig.~\ref{02}.

\begin{figure}[tbp]
\centering
\includegraphics[width=0.7\textwidth]{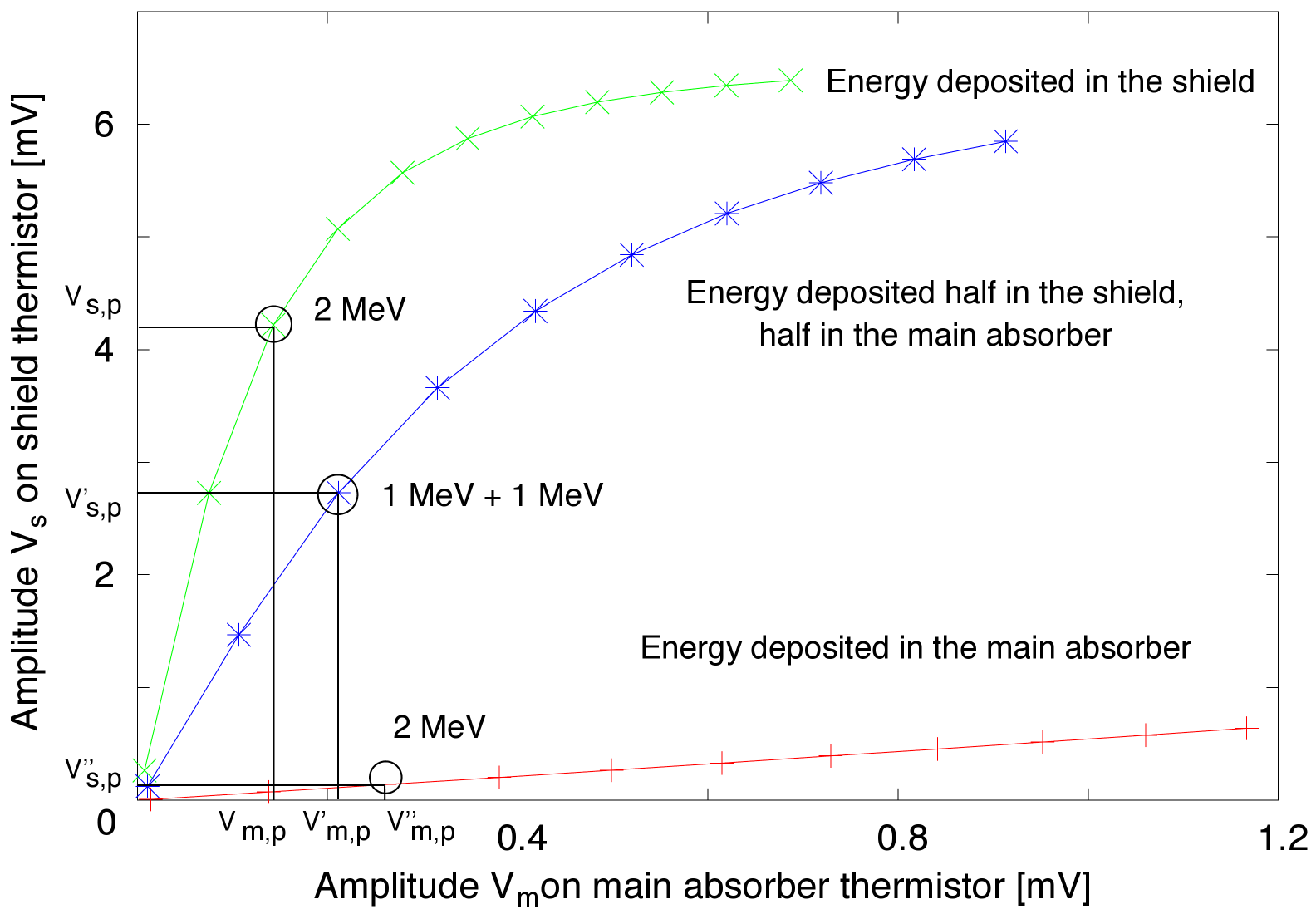}
\caption{Relation between peak values of the voltage pulses from main absorber and from the shield. The upper curve results from energy depositions in the shield only, the lower one from energy depositions in the main absorber. The intermediate curve represents events due to simultaneous energy depositions of equal energy both in the shield and in the main absorber. The energy range of this simulation is from 100 keV to 9 MeV. As an example, points corresponding to an overall energy deposition of 2 MeV are circled. In all cases, a high-energy saturation of the curves is apparent, due to the non-linearity induced by the large temperature signals $\Delta T$. This effect is not expected in the experimental results here described, since in this case the much higher operation temperature ($\sim$25~mK versus $\sim$10~mK) leads to much smaller $\Delta T / T$.}
\label{03}
\end{figure}

Information on the site of an energy deposition is given by comparing the voltage pulse amplitude read by the thermistor on the main absorber with the simultaneous voltage pulse amplitude read by the thermistor on the active shield. One can then draw a scatter plot of shield pulse amplitudes on the y-axis versus simultaneous main pulse amplitudes on the x-axis. The relation between the amplitudes is shown by the curves in Fig.~\ref{03}. The simulation shows different classes of pulses: the upper curve results from energy depositions inside the shield only, the lower one from energy depositions inside the main absorber. The intermediate curve represents events due to simultaneous energy depositions of equal energy both in the shield and in the main absorber (mixed events). The shielding element can reach saturation in the thermistor due to the low heat capacity of the shield. This simulation shows that in principle the origin of energy depositions can be determined through a scatter plot analysis.

Simulations also show the possibility of discriminating between shield and main-absorber events using the information from the shield thermistor alone, without the need for a separate thermistor on the main absorber. Energy depositions in the shield give rise to a higher and faster signal than those happening in the bulk. 

\section{Design and realization of the experimental setup}
\label{sec:setup}

Three prototype SSBs were assembled and tested. All prototypes used TeO$_2$ crystals as the main absorber. Three different materials, Ge, Si and TeO$_2$, were used as shield absorbers. The scope of this multiple choice is to show experimetally that the rejection of surface events can be efficiently achieved irrespectively of the shield composition, providing more degrees of freedom in the design of future devices. The three chosen materials have pros and cons, below summarized.

\subsection{The choice of the material for the shields}
\label{sec:material}

The advantage of Ge and Si is related to the very well developed crystallization technology. Large mass crystals can be grown, and the achievement of thin wafers from them is a routine operation in semiconductor industry. In addition, sophisticated methods for the treatment of the surfaces (grinding, polishing, etching), particularly relevant in this context, are available and spreadly used. From the bolometric point of view, Ge is a commonly used absorber with high performance~\cite{EDELWEISS}. Si is also successfully employed in bolometric experiments~\cite{CDMS}, and, at least on the paper, it should provide higher and faster pulses than Ge in equal conditions, due to its higher Debye temperature (645~K versus 360~K). In practice, both materials are excellent for the operation of low temperature macro-calorimeters, and Ge is better suited to the use of NTD Ge sensors since the mechanical coupling between equal-material elements does not pose particular problems. From the point of view of intrinsic purity, Ge is by far the purest material achievable, thanks to the zone-refining crystallization method. This reflects also on the level of the attainable radiopurity, which is excellent and well below 10$^{-12}$ g/g in U and Th concentration. Furthermore, the concentration of radioisotopes is well known thanks to the high sensitive Ge-based $0\nu\beta\beta$ experiments performed so far or currently in operation (IGEX, Heidelberg-Moscow and GERDA~\cite{rev_DBD}). The intrinsic radioactivity of Si is not known with the same accuracy, however nothing prevents in principle from obtaining similar low levels. In addition, high purity Si can be produced at prices significantly lower. This element is relevant when planning experiments with hundreds of channels, as in the present case.

However, a significant disadvantage of these two materials in the application here discussed has to be underlined. Both Ge and Si present isotropic thermal contraction due to their crystalline structure. On the contrary, the TeO$_2$ main absorber has equal thermal expansion coefficient along directions orthogonal to the $[001]$ growth axis, but a different (and furthermore negative) one along it. This means that the mechanical coupling of a thin Ge or Si element, which is anyway difficult at surfaces orthogonal to the $[001]$ axis because of the different expansion coefficients, becomes even problematic at those surfaces which are parallel to the growth axis. We experimented this problem in the present work, often observing a spontaneous detaching of the Ge or Si shields after thermal cycling. If these materials were chosen for the final application, a specific R\&D activity has to be considered to design safely the mechanical and thermal coupling of the shields.

The advantage of the TeO$_2$ shields is of course immediately related to the main (and only) disadvantage of the Ge and Si ones: they match exactly the thermal contraction features of the main absorber, after taking care of producing two types of shields (respectively orthogonal and parallel to the growth axis) and of coupling them to the appropriately corresponding main-absorber surfaces. From the point of view of the bulk radiopurity, we can consider crystalline TeO$_2$ as a safe and well known material (although not at the level of Ge and Si), thanks to the successfull $0\nu\beta\beta$ experiments performed up to now~\cite{rev_DBD,Cuoricino} and to the extensive study conducted by the CUORE collaboration on this subject~\cite{DAF}. The adequate level of the bolometric performance of TeO$_2$ is of course out of discussion. The issue here is related to the production of thin large-surface slabs with an appropriate sawing method, capable to introduce negligible surface radioactivity. Discussions with the TeO$_2$ crystal company selected by the CUORE collaboration\footnote{SICCAS, Shanghai Institute of Ceramics - Chinese Academy of Sciences - Shanghai, China} and a preliminary production show that this operation is indeed possible. The fragility of the thin TeO$_2$ shields could also be a problem for an easy mechanical assembly of the detectors, but the experience gathered in this work is reassuring under this aspect.

\subsection{Description of the detectors}
\label{sec:detectors}

\begin{figure}[tbp]
\centering
\includegraphics[width=1.0\textwidth]{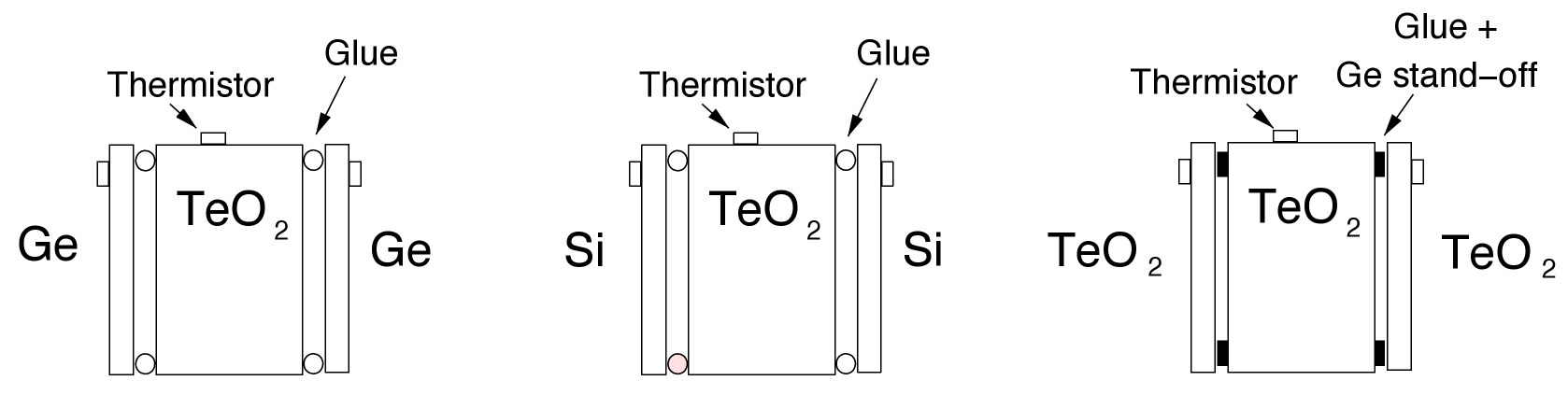}
\caption{Schematic of the three SSB prototypes.  Each detector consists of one main absorber and two active shields. The active shields are made of Ge, Si and TeO$_2$ respectively.  }
\label{06}
\end{figure}

\begin{figure}[tbp]
\centering
\includegraphics[width=0.75\textwidth]{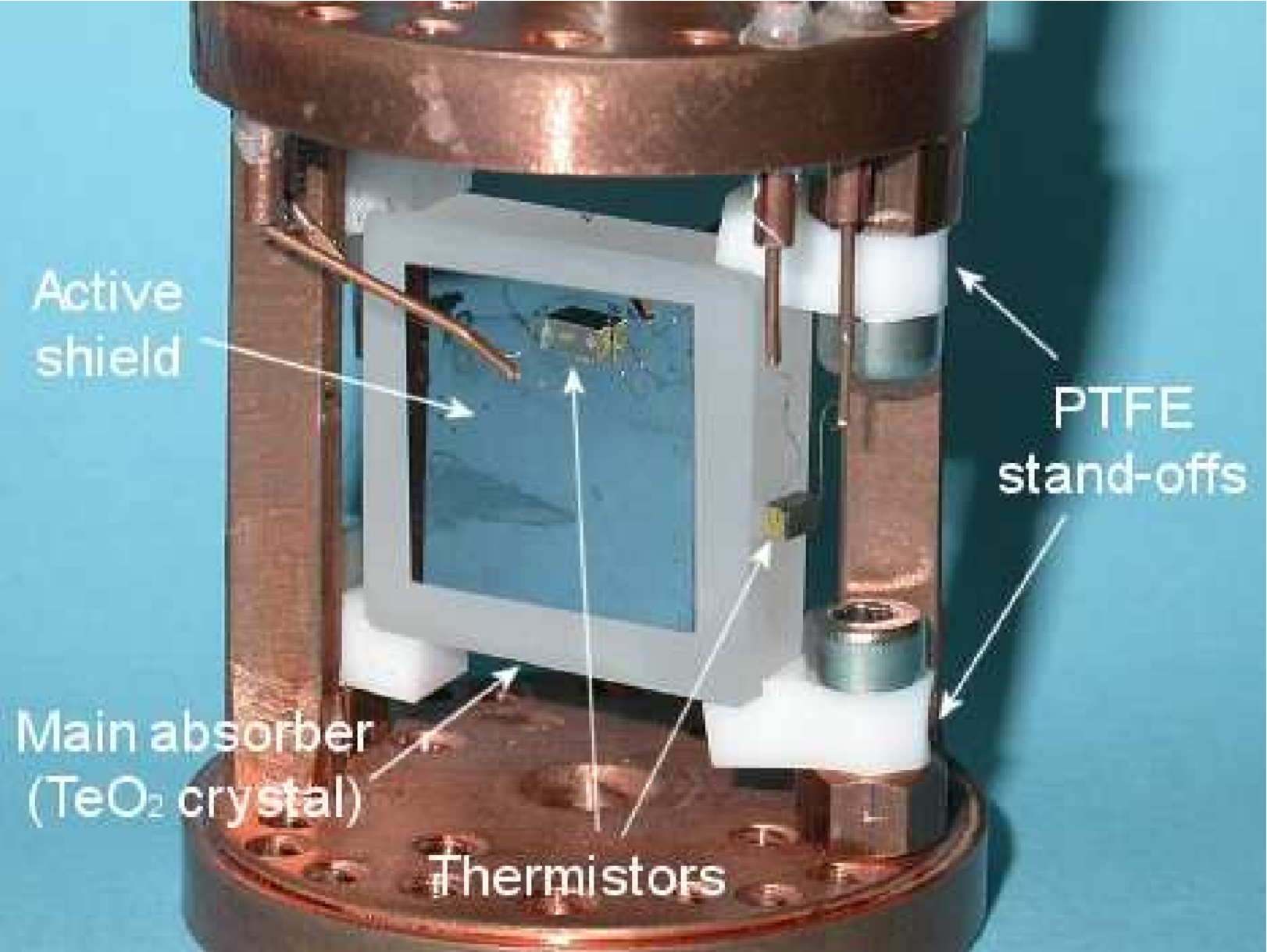}
\caption{Photograph of a surface sensitive bolometer made of TeO$_2$ and Si active shields. The light grey rectangular box is the main absorber (20$\times$20$\times$~5~mm), held by four PTFE supports 5~mm wide. One of the two Si shields is visible (the other one is on the opposite face of the TeO$_2$ crystal). The Si shield is glued at the 20$\times$20~mm face of the main absorber and appears as a square reflective thin layer with a 15$\times$15~mm surface area.}
\label{07}
\end{figure}

Fig.~\ref{06} shows schematics of the three detectors and Table~\ref{tab:03} shows their operating properties. Fig.~\ref{07} shows a photograph of the Si SSB. Each main absorber was sandwiched between two active shields, and the main absorber is thermally and mechanically coupled to the heat bath by four PTFE supports. The tests described here were conducted in the Cryogenic Laboratory at Universit\`a dell'Insubria at Como, Italy. 

\begin{table*}[h]
\caption{Main properties of the three surface sensitive bolometers used for the experimental tests: active shield material, volume and mass of the main absorber, volume, mass and heat capacity (computed at 25 mK) of the shield, thermistor volume, and resistance-temperature VRH parameters of the thermistor - see Eq.~(\ref{eq:04}).}
\label{tab:03} 
\begin{center}
\begin{tabular}{|c|c c c c c c c c|}
\hline 
Shield & V$_{main}$ & M$_{main}$ & V$_{shield}$ & M$_{shield}$ & C$_{shield}$ & V$_{thermistor}$ & $R_0$ & $T_0$ \\
material & [mm$^3$] & [g] & [mm$^3$] & [g] & [J/K] & [mm$^3$] &[$\Omega$] & [K]\\
\hline
\hline
Ge & 20$\times$20$\times$5 & 12 & 15$\times$15$\times$0.5 & 0.60 & 5.4$\times 10^{-12}$ & 3$\times$1.5$\times$1 & 2.7 & 7.8 \\
\hline
Si & 20$\times$20$\times$5 & 12 & 15$\times$15$\times$0.3 & 0.15 & 6.3$\times 10^{-13}$ & 3$\times$1.5$\times$1 & 2.7 & 7.8 \\
\hline
TeO$_2$ & 20$\times$20$\times$8 & 18 & 20$\times$20$\times$0.5 & 1.2 & 7.0$\times 10^{-11}$ & 3$\times$1.5$\times$1 & 3.1 & 5.8 \\
\hline
\end{tabular}
\end{center}
\end{table*}

In order to avoid detector saturation from cosmic rays, $2\times2\times0.5$~cm$^3$ and $2\times2\times0.8$~cm$^3$ main absorbers (see Table~\ref{tab:03}) were used instead of the $5\times5\times5$~cm$^3$ cubes used in Cuoricino. Cuoricino is located underground at Laboratori Nazionali del Gran Sasso (LNGS), Italy, and is shielded from cosmic rays under 3500~m.w.e. of rocks, while the laboratory at Insubria is aboveground.

The shields are 0.5~mm thick for Ge and TeO$_2$ and 0.3~mm for Si, and thermally coupled to the TeO$_2$ main absorber by four epoxy beads of 1 mm diameter and 50 $\mu$m thickness. The mechanical mounting is designed to minimize the risk of the shields detaching due to the difference in thermal contractions at cryogenic temperatures. This is especially true for the Ge and Si, as discussed above. An assembly method involving relatively thick epoxy beads was chosen in the Ge and Si case because the glue disks proved to be sufficiently soft to accomodate the differential material contractions. The shields were glued at surfaces orthogonal to the $[001]$ axis, for which the stress induced by the thermal contractions is less critical. 

For TeO$_2$ shields, the $[001]$~axis of the main and shield absorbers must be aligned to match the thermal contraction when cooling to low temperatures, as discussed in the previous subsection. In this case, where no difference is expected between the thermal expansion coefficients of shields and main absorber, we adopted a different gluing method, which implies a more rigid adhesion. Four Ge stand-offs (1 mm$^2$ in area and 50~$\mu$m thick) were inserted between the TeO$_2$ shields and the main absorber, with thermal coupling established by the same epoxy used in the Si and Ge case, but this time in the form of a thin veil. These stand-offs were introduced to provide a more reproducible thermal connection between the shields and main absorber, since the contact area is precisely defined by the stand-off geometry. As for Si and Ge shields, the amount of glue was carefully controlled. It was in this case much less, due the largely inferior thickness of the epoxy layer. We expect that the thermal conductance through the stand-offs, never measured, is of the same order of magnitude as that of the beads, since the interface glue-crystal should provide the dominant contribution.

The main absorber and both shields are thermally coupled to their own NTD Ge thermistor.  The thermal coupling between absorbers and thermistors is made by six epoxy beads (0.5~mm diameter and 50~$\mu$m thickness) for the main absorbers and Ge and Si shields and a single larger bead for the TeO$_2$ shields. When six beads are used, they are placed on the $3 \times 1.5$~mm~$^2$ surface of the thermistor (see Table~\ref{tab:03} for thermistor size). 

An external source of $\alpha$-particles was used to test the SSB. The $\alpha$ source was a piece of copper strip implanted with $^{224}$Ra. $^{224}$Ra emits $\alpha$-particles with a half life of 3.66~days in equilibrium with its $\alpha$ and $\beta$ emitting daughters. The main $\alpha$ lines are at 5.68, 6.29, 6.78 and 8.78~MeV. Two other weak lines sum up to 6.06~MeV in detectors with moderate energy resolution. A $\beta$-electron is emitted on average of 0.3~$\mu$s before the 8.78 MeV $\alpha$-particle. The copper strip is attached to the internal surface of a copper cylinder (not present in Fig.~\ref{07} in order to make the detector visible) which fully surrounds the detector and serves as a blackbody radiation shield. The internal diameter of the cylinder is equal to one of the two copper disks visible in Fig.~\ref{07}, above and below the detector. The copper strip, with an area of about 1~cm$^2$, is placed at the center of the shield. Since the surface area of the main absorber is larger than that of the shields in two cases (see Table~\ref{tab:03}), we expect that a fraction of the $\alpha$-particles will hit the absorber directly. The $\beta$-electrons, due to their much longer range, can penetrate the shield and deposit energy both in the shield and in the main absorber. In the former case, the $\alpha$ events are recognized as main-absorber events, while in the latter the $\beta$ events are mixed events.

The detectors were cooled down separately in a low power dilution refrigerator capable of reaching a base temperature of $\sim$20~mK. The detector thermistors were DC biased through a voltage supply and a load resistor at room temperature between 2~G$\Omega$ to 20~G$\Omega$, chosen according to the specific sensor characteristics. The typical operating temperature was $\sim$25~mK, corresponding to a thermistor resistance of $\sim$10~M$\Omega$. The thermistors used in these tests were optimized for temperatures higher than those assumed in the simulations of Section~\ref{sec:model}. The VRH parameters are in this case $R_0 =$~3.1~$\Omega$ and $T_0 =$~5.8~K. The static bias voltage across the thermistors ranged typically from 10 to 20~mV. Voltage pulses were read out by a DC-coupled low noise differential voltage amplifier, followed by a filtering single-ended stage. The front end electronics was at room temperature. The signals were acquired by a 12-bit transient recorder, collecting 1024 points for each pulse, and registered for off-line analysis. The main purpose of the experiment was to verify and understand the surface event discrimination capabilities of the detector and energy resolution was not optimized. The data sets discussed in the next section were acquired during measurements lasting typically several hours.

\section{Experimental results and data discussion}
\label{sec:results}

The thermal model described in Section~\ref{sec:model} for SSBs was used to simulate the detector behavior assuming the Cuoricino thermal parameters. In case of energy absorbed in the shield (surface event), it predicts a substantial shape and amplitude difference between the signal read out by the shield thermistor and that read out by the main-absorber thermistor (see Fig.~\ref{02}c). A fast and high signal in the shield thermistor is accompanied by a simultaneous low and slow signal in the main-absorber thermistor. This is the main ingredient which allows the identification of the event origin. The first objective of the experimental tests is to verify this behavior, at least qualitatively.

\begin{figure}[tbp]
\centering
\includegraphics[width=0.8\textwidth]{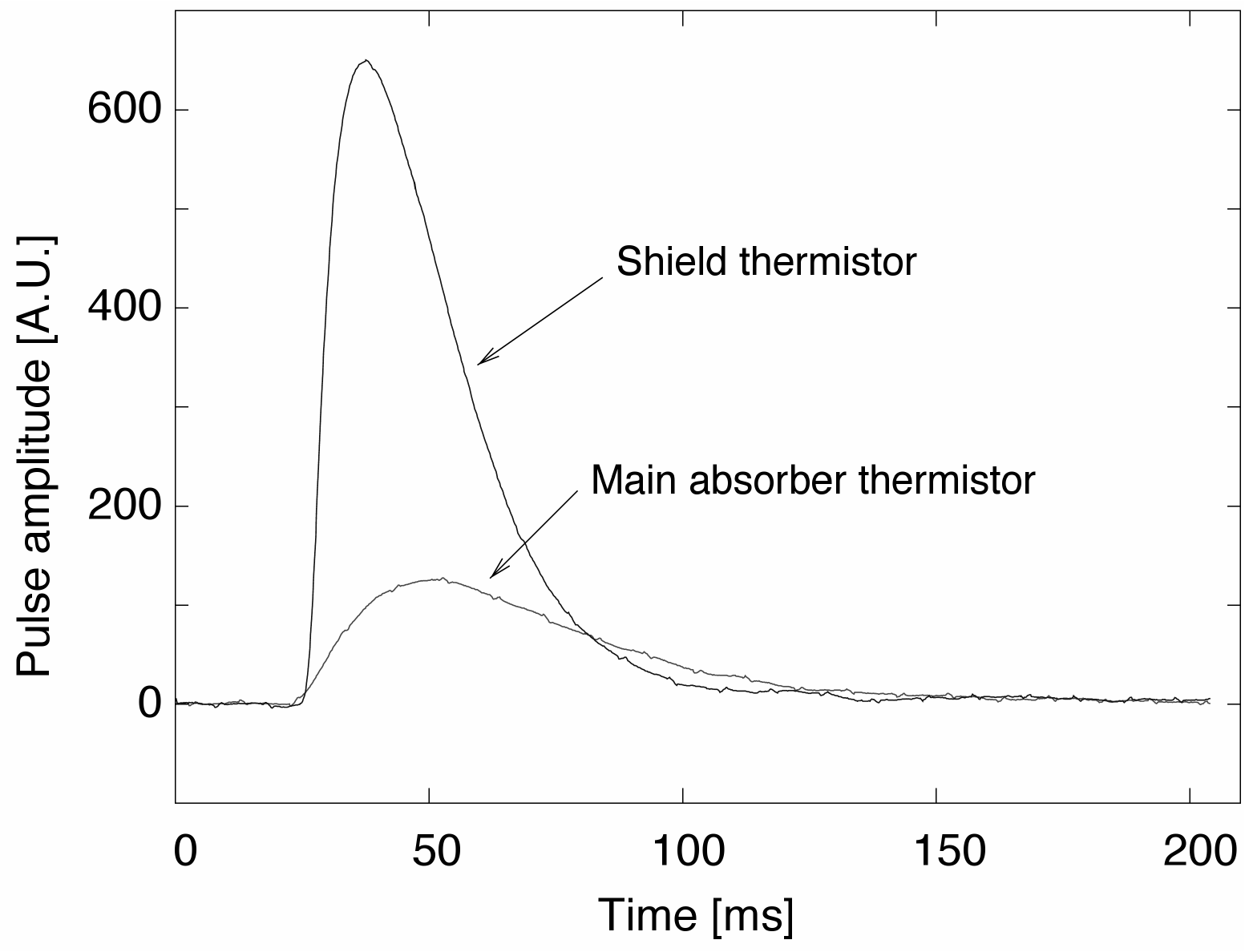}
\caption{Experimental pulses generated by an alpha particle fully absorbed in the Ge active shield.}
\label{08}
\end{figure}

We have not simulated directly the detectors realized here because ad hoc measurement of the thermal parameters have beeen performed only on Cuoricino detectors, and not for the type of mounting used in the present set-up. For example, the thermal coupling between main absorber and heat sink is different here from that adopted in Cuoricino, and has never been measured. In addition, the heat capacities and operating temperatures for our set-up are very different from those used in Cuoricino. Neverthelss, the basic detector structure is analogous to that simulated, and therefore we expect a qualitative agreement between simulations and experimental data. This agreement can be clearly appreciated in Fig.~\ref{08}, which shows experimental pulses from both thermistors originated by an $\alpha$-particle fully absorbed in one of the Ge shields of the SSB. The difference between the two pulses is attenuated by a low pass filter used in the experiment with a cutoff frequency immediately below 50 Hz, and limiting therefore the pulse rise time to about 6~ms. This filter helps in improving the signal to noise ratio without affecting the discrimination capability of the realized SSBs. The pulses acquired with the SSBs provided with Ge, Si and TeO$_2$ shields exhibit similar features. This is not surprising, since the heat capacity of the shield thermistor is of the order of 10$^{-10}$~J/K at 25~mK (see Table~\ref{tab:02}), therefore dominating over the heat capacities of the shields independently of the material (see Table~\ref{tab:03}). Moreover, the thermal conductances beteween shields and main absorber are similar, as discussed in Section~\ref{sec:detectors}. However, this parameter is difficult to reproduce and is probably responsible for the observed differences.

\subsection{Event origin identification through pulse-amplitude scatter plots}
\label{sec:scaplo}

\begin{figure}[tbp]
\centering
\includegraphics[width=1.0\textwidth]{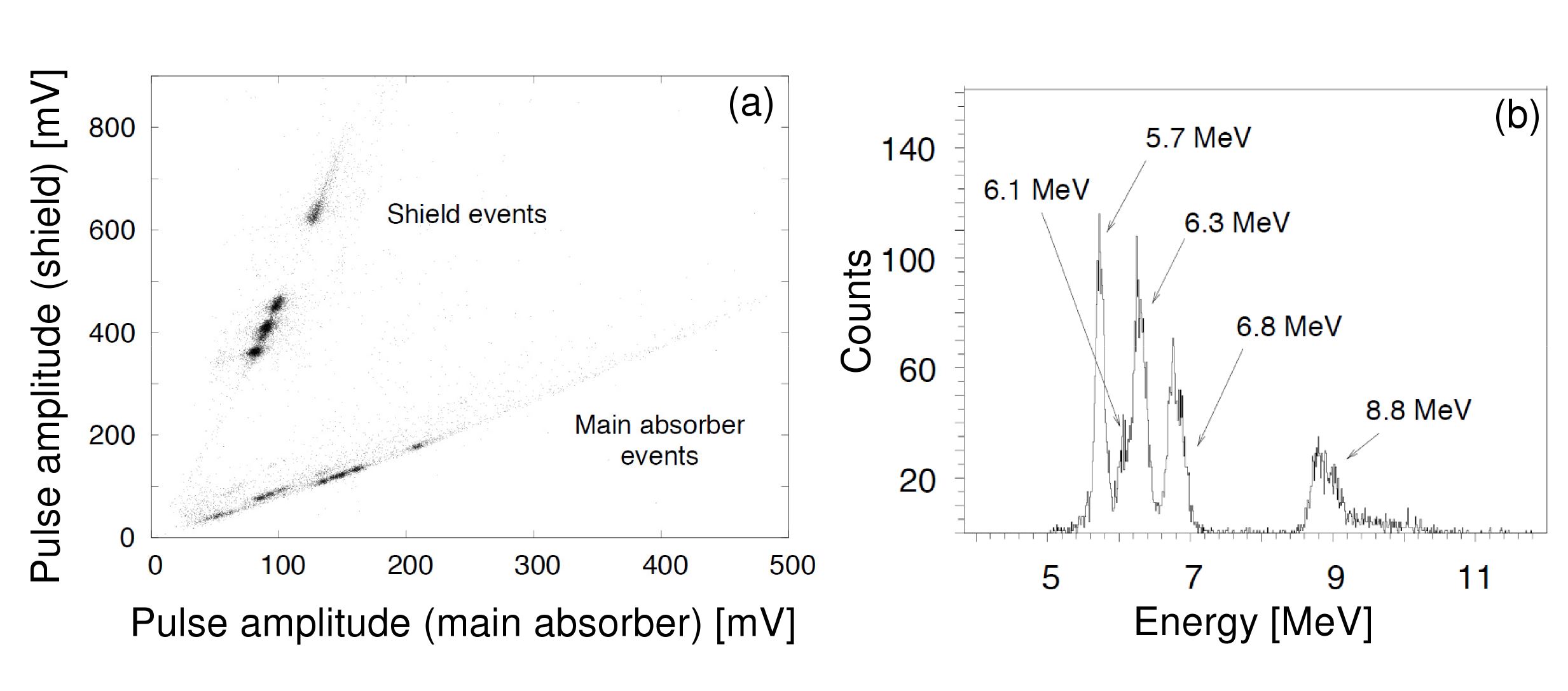}
\caption{(a) Experimental scatter plot showing the relationship between main-absorber pulse amplitudes and shield pulse amplitudes acquired for one of the Ge shields. Two populations of events appear, distributed in bands with different slopes. These bands include mainly events from only one of the absorbers. (b) Energy spectrum from the shield read-out: the higher slope band in the scatter plot is selected so that only pulses due to total energy depositions in the shield are kept. The expected structure of the spectrum of the source, which is outside the detector and faces the shield, is clearly extracted from the raw spectrum after surface event discrimination.}
\label{Ge-slab}
\end{figure}

\begin{figure}[tbp]
\centering
\includegraphics[width=0.8\textwidth]{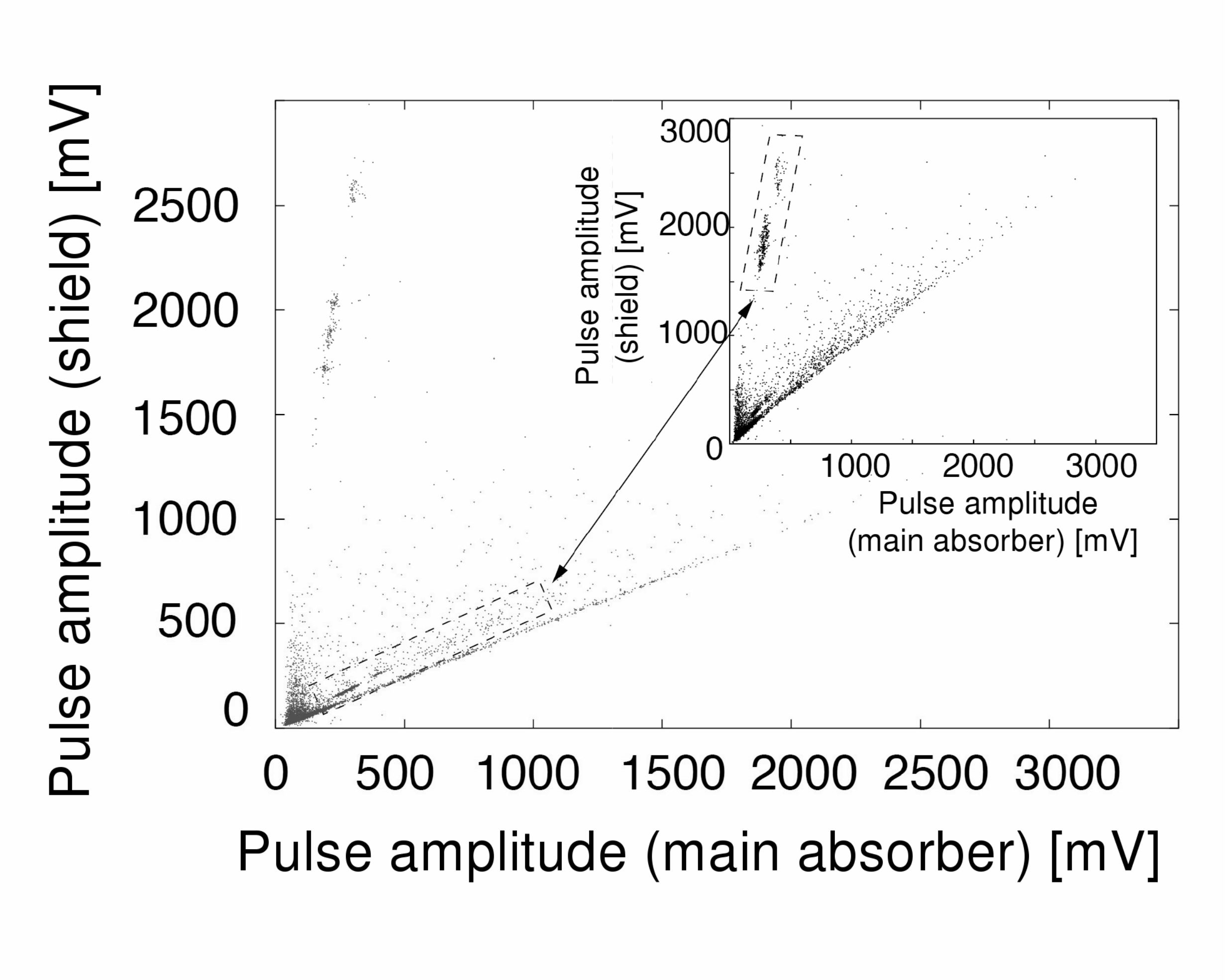}
\caption{Experimental scatter plot reporting shield pulse amplitudes vs. main-absorber pulse amplitudes for the first TeO$_2$ shield. In the inset, the same for the second TeO$_2$ shield. Surface events due to energy depositions on one shield can be reciprocally seen on the scatter plot corresponding to amplitudes seen by the thermometer on the other shield.}
\label{Te-slab-rec}
\end{figure}

In the scatter plot obtained from the Ge-shield detector and shown in Fig.~\ref{Ge-slab}a, energy depositions in the shield can easily be distinguished from those in the main absorber since they are distributed in a band with a steeper slope. The main-absorber event band exhibits a fine structure with a sub-band characterized by a slightly higher slope. This effect is due to alpha particles impinging on the shield which is not read out. These events produce a thermal response which is slightly different from that of those occurring in the main absorber, and therefore they appear as a separate population. This effect is explained in more detail for the detector with TeO$_2$ shields (see Fig.~\ref{Te-slab-rec} and the related explanation in the text). 

The scatter plot can be compared with the simulation of Fig.~\ref{03}. The qualitative agreement between experiment and simulation is once again evident. The simulation shows a shield pulse amplitude saturation which does not appear in the experimental data. The lower temperatures used in Cuoricino implies much lower heat capacities with respect to the experimental case. This determines a non-linearity of the pulse amplitude-energy relation when particles impact the shield, while in the experimental data this relation is linear both for the shield and the main-absorber events, since the detectors are operated at 25 mK. This explains why the two main event populations in Fig.~\ref{Ge-slab}a are distributed along straight lines. 

Bulk events giving pulses in the lower band in Fig.~\ref{Ge-slab}a are mainly due to natural $\gamma$ radioactivity and to cosmic-ray muons passing through the main absorber without interacting with one of the shields. The points between the two bands are mainly due to particles which deposit energy in both main absorber and shield. Most of these mixed events are due to muons crossing both the read out detector elements, as shown in a previous work~\cite{A12}, and to $\gamma$'s undergoing a Compton interaction in one of the two elements followed by a Compton interaction or photoelectric absorption in the other one. As mentioned in Section~\ref{sec:model}, the possibility of identifying mixed events is important for experiments such as CUORE, as this will add the capability of discriminating energy depositions due to TeO$_2$ surface contamination, although this contribution looks less dangereous than that provided by the inert Cu structure. 

The effectiveness of the discrimination power of this technique is shown in Fig.~\ref{Ge-slab}b. The structure of the external $\alpha$-source is clearly extracted from the raw energy spectrum of the shield after selecting the higher slope band shown in fig.~\ref{Ge-slab}a.

Data analysis of the TeO$_2$ SSB evidenced the possibility to observe energy deposition in one of the active shields by reading the thermistor on the other shield.
Fig.~\ref{Te-slab-rec} shows the scatter plot obtained from the individual reading of the thermistors on one shield and on the main absorber, while acquiring signals not just from these two thermistors, but also from the sensor on the second shield. In addition to the usual main-absorber and shield event bands, a second population appears in a band with a slope slightly higher than that of the main-absorber band. To identify the origin of this population, we selected the shield events band in the scatter plot obtained from the thermistor on the other TeO$_2$ shield, shown in the inset, and cut on the first graph. The result of such selection is that the additional population close to the bulk region identifies pulses due to $\alpha$ energy depositions on the other shield. In the shield thermistor read-out, we would, in principle, expect no appreciable difference between pulses corresponding to energy releases in the other shield or in the main absorber. However, the different heat flow paths in the two cases (an additional thermal impedence is present for energy deposited in the shield) lead to slight differences in pulse shapes (see Section~\ref{sec:dt}). Since the pulse amplitudes in the scatter plots are determined through a digital optimum filter which uses an average pulse of bulk events as signal template, the amplitude reconstruction depends on the pulse shape. This determines the additional population observed in the scatter plot.

\begin{figure}[tbp]
\centering
\includegraphics[width=0.8\textwidth]{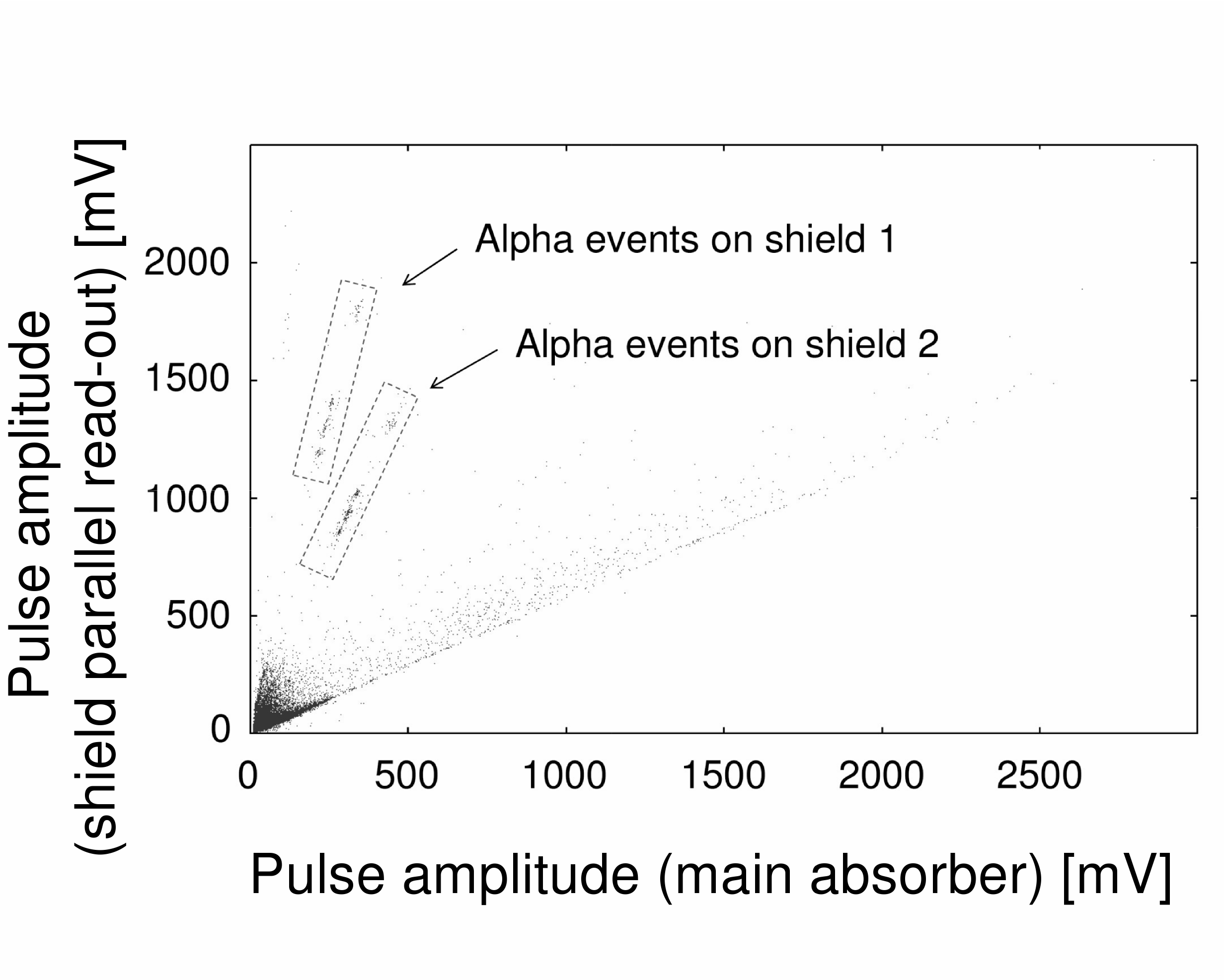}
\caption{Experimental scatter plot for TeO$_2$ SSB corresponding to a parallel read-out of the shield thermistors. The two boxes contain events observed in the two shields}\label{scat-paral}
\end{figure}

The signals from the shields can be acquired by connecting the thermistors in parallel. This interesting possibility would reduce significantly the number of read-out channels, which is a critical parameter for cryogenic systems that have to reach temperatures below 10~mK. In fact, each read out wire is a source of power load. This possibility was tested with Si and TeO$_2$ SSBs, and the results from the latter are shown in Fig.~\ref{scat-paral}.  The two boxes identify $\alpha$ interactions on each shield. The difference in the slope of the bands is due to unequal thermal conductance between the two shields and the main absorber, as the same thermal coupling is difficult to reproduce experimentally. The use of the Ge stand-offs for the TeO$_2$ case (see Fig. \ref{06}) has however improved considerably the reproducibility of the couplings, making the results on the parallel readout particularly significant.

\subsection{Event origin identification through pulse shape analysis}

As mentioned in Section~\ref{sec:model}, shield events can be distinguished from main-absorber events by a pulse-shape analysis of the signal from a shield thermistor, without additional information from the main thermistor. This discrimination possibility was confrimed in all the three detectors realized, by analyzing in particular the rise time of the pulses from the shield thermistors.

\begin{figure}[tbp]
\centering
\includegraphics[width=1.0\textwidth]{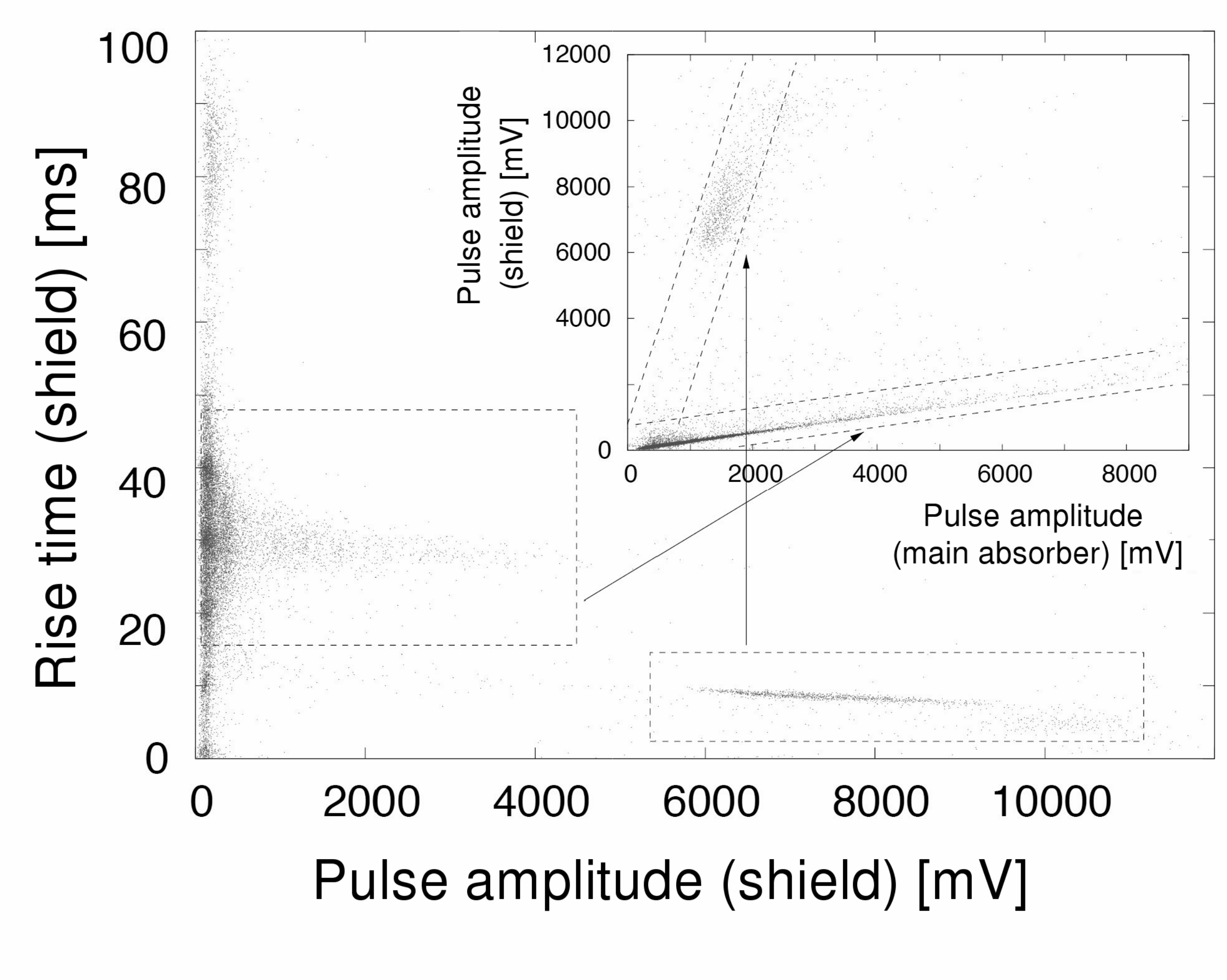}
\caption{Shield pulse rise-time vs. amplitude for the detector with Si shields. Two classes of events corresponding to different $\tau_r$ can be identified. In the inset, scatter plot showing shield-pulse vs. main-absorber pulse amplitudes for one Si active shield. The relationship between the two graphs is indicated with arrows: pulses corresponding to shield energy depositions can be distinguished because of lower $\tau_r$ with respect to main-absorber energy depositions.}
\label{rise-scat}
\end{figure}

In order to show the potential of this method, we report the results obtained with the detector equipped with Si shields as an example. Fig.~\ref{rise-scat} shows the rise time $\tau_r$ vs.~pulse amplitude for one shield thermistor of the Si SSB. This can be compared to the simulated behavior described in the last part of Section~\ref{sec:model}. The Si shield sees two clearly recognizable groups of pulse amplitudes: fast pulses due to particle interactions in the shield ($\tau_r$$\sim$~10~ms) and slow pulses originating from energy depositions in the main absorber ($\tau_r$$\sim$~30~ms). In the inset, the scatter plot shows that the class of fast pulses corresponds to shield events.

The clear event origin discrimination achievable with shield thermistors offers the possibility of reading out the entire detector using a single wire pair. The main-absorber thermistor could be eliminated and the shield thermitors could be connected in parallel at the detector level. Only one wire pair would be used to extract the common signal from the cryostat. From this signal, it would be possible to obtain the fundamental information concerning the energy deposited in the main absorber from the analysis of the slow pulses (corresponding to bulk events), while the surface events could be cut by rejecting the fast rise time pulses. Also mixed events with a substantial energy fraction deposited in the shield would be identified (se Fig.~\ref{02}(b)). Future tests will be needed to measure the effect on the energy resolution relative to the usual main-absorber thermistor, and optimize any trade-off between background rejection and energy resolution acting on the thermal parameters of the detector.

\subsection{Quantification of the surface-event rejection power}

In order to estimate the impact of the proposed method on the sensitivity of future experiments, it is useful to attempt a preliminary evaluation of the surface-event rejection power. We have performed this investigation using the identification method based on the scatter-plot analysis. The detector selected for this study is the one with TeO$_2$ shields read out in parallel (the third one in Fig.~\ref{06}), since it is the most similar to a practical device for double beta decay search. First we have fitted the points corresponding to bulk events with a straight line starting from the origin in the scatter plot reported in Fig.~\ref{scat-paral}. These points are easily identifyable and correspond to the narrow band with the smallest slope and ending at about 2500 mV on the x-axis. Let $\beta$ be the slope of this band. We can then construct a parameter, associated to each event and defined as {\em SEI} (Surface Event Index), in the following way:
\begin{equation}
{\emph SEI} \equiv { \Delta V \ [shield] \over \beta \cdot \Delta V \ [main \ absorber] } - 1
\end{equation}
where $(\Delta V \ [main \ absorber], \Delta V \ [shield])$ is a pair of values defining a point on the scatter plot and corresponding to an event registered by the detector.

For construction, {\em SEI} will be 0 (within experimental errors) for bulk events, while it will be $> 0$ for events characterized by a direct energy deposition in one of the two shields. Looking at the distribution of {\em SEI}, we expect then a peak centered on 0 and corresponding to bulk events and two peaks centered on values higher than 0 and corresponding to shield events. We expect also a small population of events lying outside the three main peaks, corresponding to the cases of mixed main absorber-shield energy depositions, already discussed in Section~\ref{sec:scaplo}.

The separation between the bulk event peak and the shield event peaks in the {\em SEI} distribution allows to evaluate the rejection power of the method. By fitting the peaks with Gaussians, it is possible to determine the leak of a Gaussian related to shield events into a region centered on 0 and corresponding to bulk events selected with a given efficiency. Of course, we expect that the rejection power depends on the threshold for the energy deposited in the main absorber. The lower the threshold, the lower the signal-to-noise ratio, with consequent merging of the distributions related to bulk and surface events.

\begin{figure}[tbp]
\centering
\includegraphics[width=1.0\textwidth]{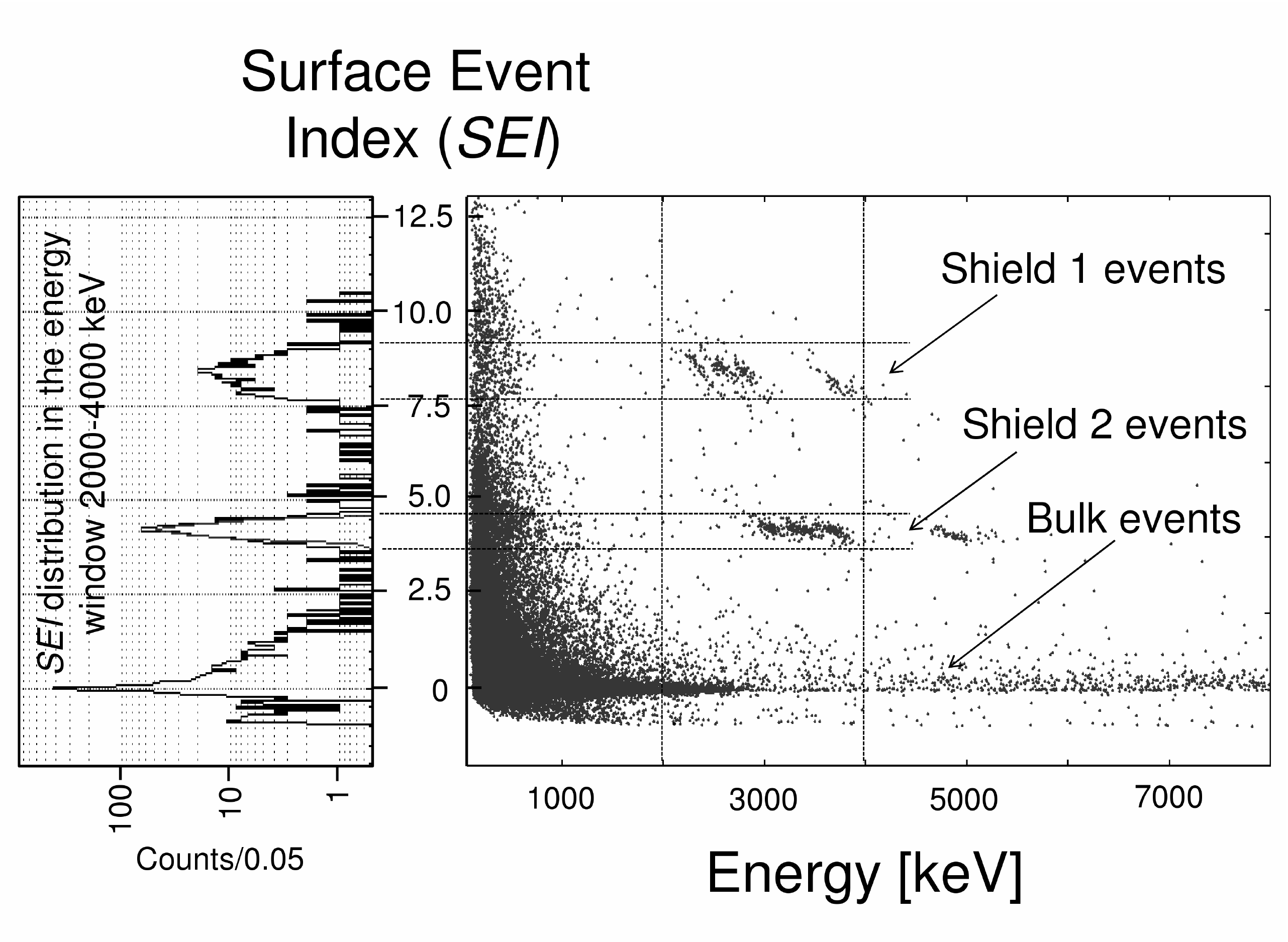}
\caption{The Surface Event Index ({\em SEI}) introduced in the text is plotted as a function of the energy deposited in the main absorber for the detector with two TeO$_2$ shields read out in parallel. The distribution of {\em SEI} is projected on the left for events in the energy window 2000-4000 keV. Three event populations (bulk, shield 1 and shield 2 events) appear clearly. The separation power of the event classes is discussed in the text.}
\label{fig:rej}
\end{figure}

The results of this method applied to the detector with two TeO$_2$ shields are appreciable in Fig.\ref{fig:rej}. In the x-axis, we report the pulse amplitude read by the bulk thermistor calibrated in energy. The histogram of these pulse amplitudes shows clearly two peaks identifiable as the $^{40}$K peak (1460 keV) and the $^{208}$Tl peak (2615 keV) due to the natural $\gamma$ radiactivity in the laboratory. This identification makes the energy calibration of the x-axis possible. On the y-axis of the main plot of Fig.\ref{fig:rej}, we report the parameter {\em SEI} constructed as described above. As expected, it is possible to observe a horizontal band of events at {\em SEI}~$\sim 0$ (bulk events), one at {\em SEI}~$\sim 4$ (shield 2 events) and one at at {\em SEI}~$\sim 8$ (shield 1 events). In the bands of shield events, the four clusters corresponding to the main lines of the $\alpha$ source used to simulate surface contamination are appreciable. The inclined shape of these clusters (more evident for events on shield 1) are probably explanable as a position effect due to the presence of point-like contacts between the shield and the main crystal provided by the Ge stand-offs. If an $\alpha$-particle deposits its energy close to a stand-off, it generates a higher pulse in the main-absorber thermistor with respect to an energy release far from the shield-crystal contact point. The effect is opposite for the shield thermistor.

After selecting an energy window (2000 - 4000 keV) containing the Q-value of $^{130}$Te Double Beta Decay (2527 keV), the three horizontal bands described above are projected into the {\em SEI} distribution showed in log scale on the left side of Fig.~\ref{fig:rej}. Three peaks appear, corresponding to bulk, shield 1 and shield 2 events. The bulk event peak exhibits a tail at high {\em SEI} values due to the mixed events discussed previously. In order to assess the rejection power, the shield 2 event peak has been fitted with a Gaussian, centered on 4.208 and with $\sigma = 0.127$. The distribution of the bulk events, including the mixed event tail, stops at about {\em SEI}~$=2.5$, about 13 $\sigma$'s away from the centroid of shield 2 events. This analysis shows that in the double beta decay region there is no contamination of $\alpha$'s impinging on shield 2 in the bulk event population and that the rejection power is essentially 100\%. A similar result is achieved when analysing the shield 1 events. These preliminary results are very encouraging and indicate clearly that a full recognition of surface events is possible with this technology.

Although the main purpose of the present method is to discriminate surface events emitted by Cu surfaces, we believe that it has the potential to reject mixed events as well, especially those really dangerous for the search for $0\nu\beta\beta$. In Fig.~\ref{fig:rej}, the mixed events are mostly concentrated close to the bulk event band. This is due to the fact that in an aboveground experiment most of the mixed events are due to energetic cosmic muons with inclined trajectories which cross both the main absorber and the shield. Due to the thinness of the shield, these muons normally deposit much more energy in the main absorber than in the shield. A fraction of them could anyway be discriminated by fitting the SEI distribution around 0 with a symmetric gaussian plus a tail on the right side. However, this operation is not relevant here since these events will not be present in an underground set-up, where the dangerous background is mainly due to $\alpha$ radioactivity. As already pointed out, in order to simulate a $0\nu\beta\beta$ $\alpha$-particles must deposit at least $\sim$1.5 MeV in the shield. The energy share in this case is such to produce a mixed event similar to that depicted in Fig.~\ref{02}(b), and the corresponding band on the scatter plot is visible in Fig.~\ref{03}. Therefore, after the selection of an energy interval around the $0\nu\beta\beta$ energy, we expect a SEI distribution with a clear gap between the bulk event peak and the mixed event population. In order to quantify the rejection power in this case, we are designing an experiment, to be performed underground, in which a superficial $\alpha$ source is intentionally deposited on the main-absorber surface, below a TeO$_2$ shield.   

\section{Future prospects: Pulse Shape Discrimination with main-absorber signals}
\label{sec:dt} 

The prototype SSBs for discriminating surface contamination events described here consist of three separate thermistor read-outs, one for the main absorber and one for each of the two active shields. Data analysis and simulations show an interesting possibility of doing this discrimination using only the thermistor connected to the main absorber. This approach obeys to a completely different philosophy with respect to the results presented in the previous Section, since the concept of auxiliary bolometers is abandoned and the shields play a passive role aiming at the modification of the signal shape. This method looks very promising for real application in future experiments, and will be discussed with some detail in this Section.

\begin{figure}[tbp]
\centering
\includegraphics[width=1.0\textwidth]{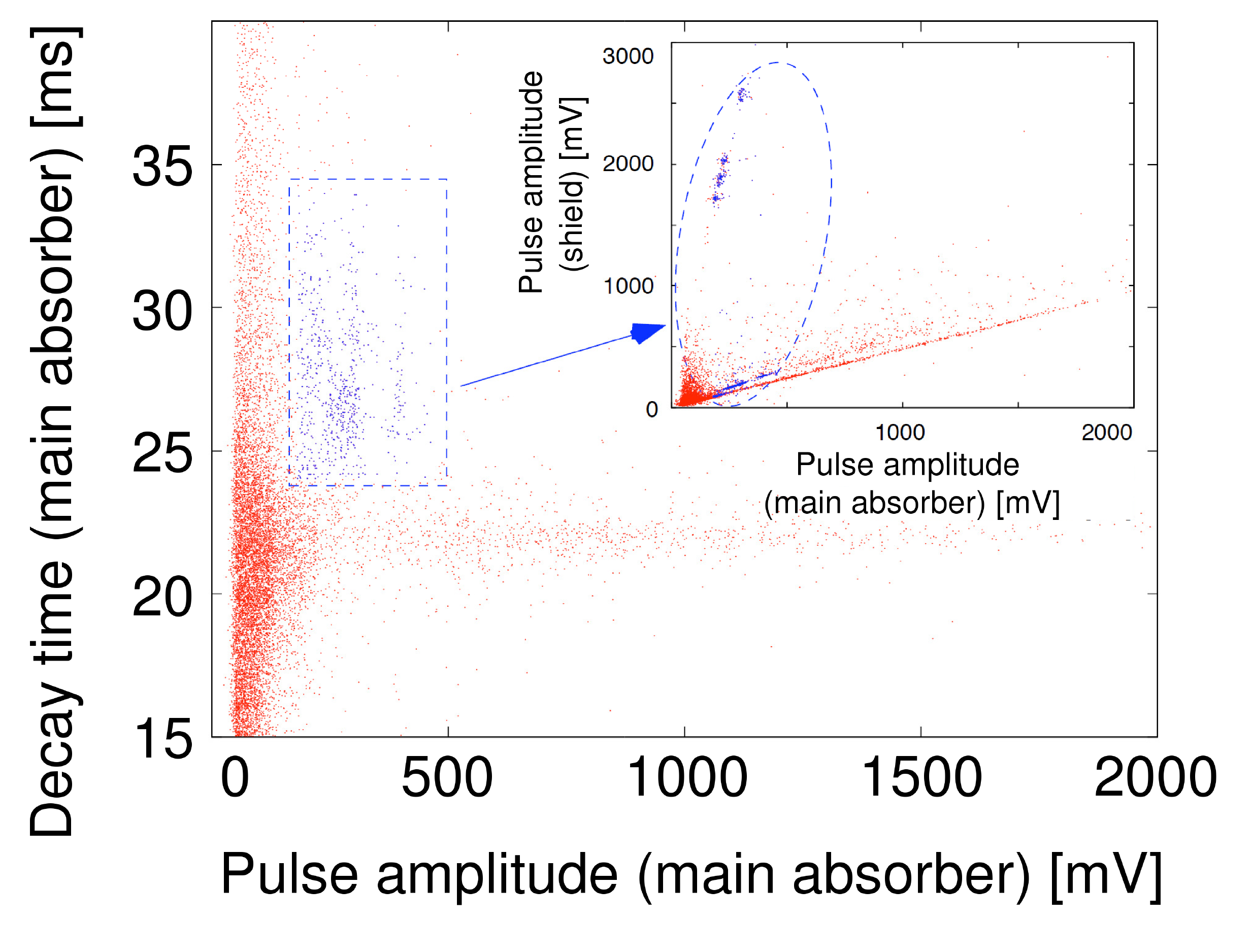}
\caption{Main-absorber pulse decay time vs. pulse amplitude for the detector with TeO$_2$ shields. A structure with groups of events with higher $\tau_d$ can be isolated from the region that identifies usual bulk events,characterized by $\sim$~22 ms decay time rather independent of the amplitude. In the inset, scatter plot of shield-pulse vs. main-absorber pulse amplitudes using one TeO$_2$ shield. As can be seen in the two graphs, events in the shields can be separated from the main absorber events by selecting events with high $\tau_d$.}
\label{decay-scat}
\end{figure}

Fig.~\ref{decay-scat} shows the distribution of decay time (90\%-30\%) vs.~amplitude for pulses from the main absorber of the TeO$_2$ SSB. The plot shows that the typical decay time is of the order of 22~ms, rather independent of amplitude above a given threshold. The spread of the decay times at small amplitudes is an effect ascribable to noise and especially to 'pile-up' pulses. These latter are a class of partially-overlapping pulses (in the time domain) whose tricky superposition leads them to be recognized as a single pulse with higher decay time. The aboveground operation of the detector and the intrinsic slowness of bolometers make the pile-up effect particularly severe. The plot shows also different groups of pulse amplitudes corresponding to longer decay times. These groups are included in the box outlined by dotted lines and correspond to the points circled in the inset, which shows the familiar scatter plot of shield pulse amplitudes vs. main-absorber pulse amplitudes. This shows clearly that shield events are identifiable as longer decay time pulses when read out by the thermistor on the main absorber. The additional population selected close to the bulk event band is due to events occuring in the other shield and discussed previously (see Fig.~\ref{Te-slab-rec} and related discussion). The event discrimination in Fig.~\ref{decay-scat} using decay times is not as clear as in the inset, however further optimization of the thermal parameters may be possible in the future. 

In order to understand better this effect, the decay time as a function of the pulse amplitude was analyzed through our pulse shape simulation tool, based on the thermal model of Section \ref{sec:model}. The simulations confirm that a longer decay time is expected for events originating in the shield. This is due to the additional thermal impedance between shield and main absorber, which slows down the thermal pulses as seen by the main-absorber thermistor. An important point emerging from the simulation results is that this thermal effect is enhanced by the presence of an element with large heat capacity connected to the shield. In ordinary conditions, this element is the electron system of the shield thermistor. Another result is that the effect does not require the shield thermistor to be linked to the heat bath through the wires for signal acquisition. In principle, therefore, the thermometer can be replaced by any passive element with appropriate heat capacity with the purpose to enhance the pulse shape difference.

\begin{figure}[tbp]
\centering
\includegraphics[width=1.0\textwidth]{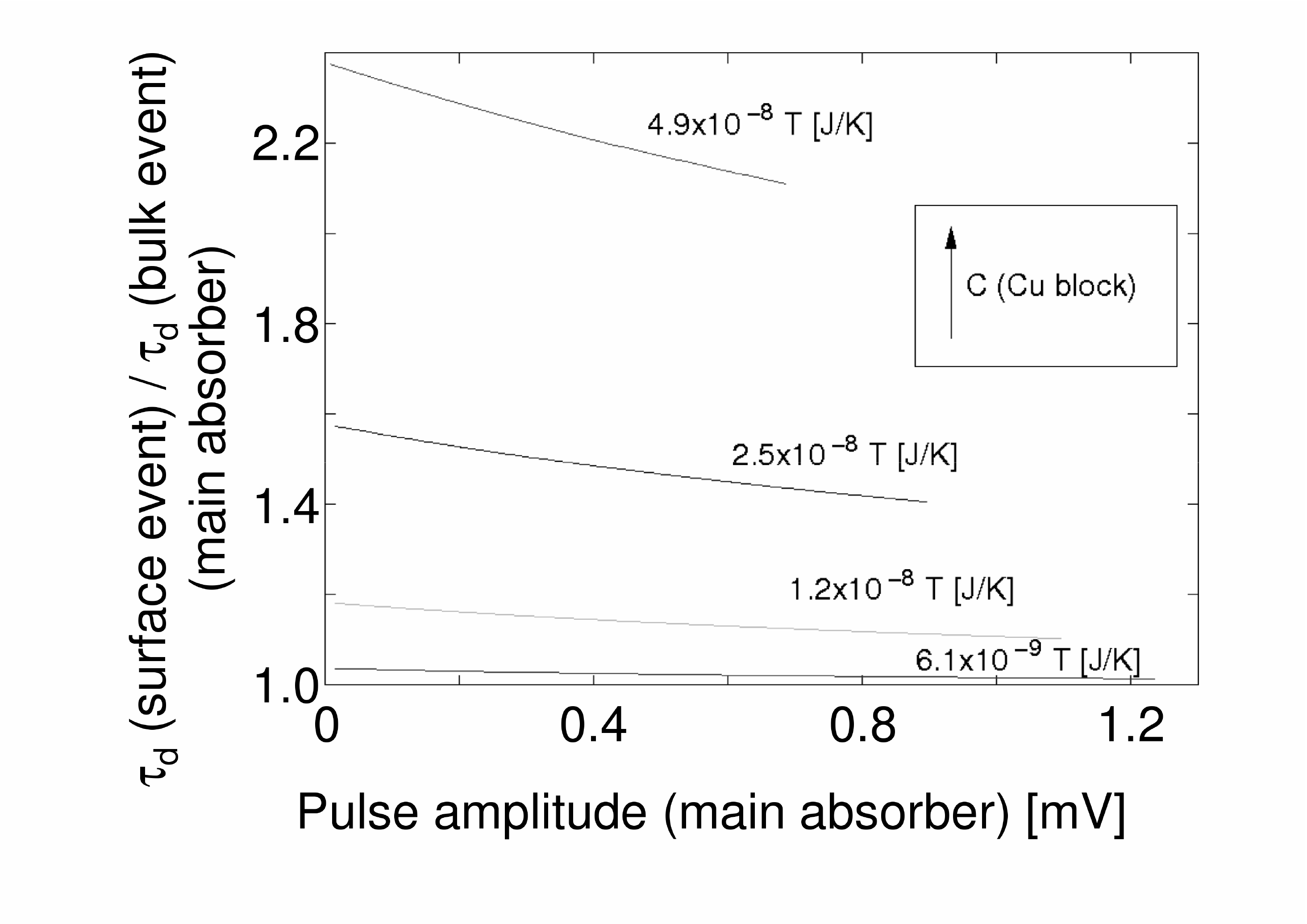}
\caption{Ratio between the decay time $\tau_d$ for main-absorber pulses corresponding to shield energy deposition and that corresponding to main-absorber energy depositions vs. main-absorber pulse amplitudes. The range of the energy deposited in the main absorber is 0-9 MeV. The simulated detector is a SSB with a TeO$_2$ cubic main absorber (5~cm side length), a single Si active shield (5$\times$5~cm$^2$
surface and 300~$\mu$m thickness) and a small Cu block with variable size (maximum volume 0.5~mm$^3$) glued on the shield. The graph shows that the $\tau_d$ ratio increases with the heat capacity of the Cu block.}
\label{decay-sim}
\end{figure}

To analyze quantitatively this possibility, a series of simulations were carried out with a 5$\times$5$\times$5~cm$^3$ TeO$_2$ cubic main absorber and a single Si shield with a small Cu block thermally attached to the Si.  In this model there is only one thermistor attached to the main absorber, and none on the shield.  The simulations estimate the decay time vs. amplitude for an energy range 100 keV - 9 MeV. The size of the Cu block was varied, spanning a heat capacity range from 6.1$\times$10$^{-9}$$\cdot$T J/K to 4.9$\times$10$^{-8}$$\cdot$T J/K, where $T$ is measured in Kelvin. The base temperature was set at 9~mK, and all other parameters listed in Table~\ref{tab:02} were used, in order to simulate Cuoricino / CUORE type detectors. The results of these simulations for several different heat capacities of the Cu block are illustrated in Fig.~\ref{decay-sim}.  It shows the ratio of the decay-time $\tau_d$ of a shield event to a main-absorber event as a function of the pulse amplitude, as observed by the main-absorber thermistor. The simulations show clearly the increasing trend of this ratio as the size of the Cu block increases. This technique is potentially a very powerful tool for large-scale experiments with a large number of detectors. It allows for shield event discrimination resulting from surface contamination thanks to pulse shape analysis techniques and without the complexity and heat load of additional thermistors and their accompanying read-out wires. 

A price to pay is a reduction of the pulse amplitude, as clear from Fig.~\ref{decay-sim} when one looks at the maximum amplitude corresponding to 9 MeV energy deposition. However, a reasonable compromise can be found. For example, the second smallest copper block (which has a heat capacity at $\sim$10 mK $\sim$200 times higher than that of the Si shield and similar to that of the thermistor) determines an acceptable pulse amplitude reduction of the order of 10\%, providing a decay time longer by the same amount. With the very high signal-to-noise ratio pulses expected in the $0\nu\beta\beta$ region, this difference in shape should be easily identified. The capability fo this method to identify also mixed events needs to be investigated. We expect that if a considerable amount of the energy is deposited in the shield the slow time constant related to the discharge of this energy into the main absorber should modify the pulse shape read by the main-absorber thermistor. An experimental confirmation is however mandatory.

In order to simplify the detector structure and to avoid the Cu block, one could choose a shield material with a high specific heat at low temperatures. This material cannot be a normal metal, for which the specific heat is proportional to the temperature and definitely too high, given the large size of the shield. One should therefore choose a dielectric or a superconductive material (in which the electron contribution to the specific heat vanishes exponentially). However, this choice is not straight-forward, as this material should satisfy many constraints simultaneously: to have a low Debye temperature in order to get sufficiently high heat capacities; to be easily put in the form of large surface single crystals; to exhibit a very low intrinsic bulk radioactivity; to have surfaces easily cleanable by polishing and / or etching; to have thermal contractions which match reasonably well the TeO$_2$ ones. This means to open a new R\&D line with uncertain outcome. Seeking simplification, another solution could be to decrease substantially the thermal conductance between shield and main absorber, in order to enhance the pulse shape difference without introducing an additional heat capacity. This is not easy too, since the mechanical fixing of the shield requires three -- four contact points with surface of the order of a square millimeter (a smaller area would not provide a safe attachment). Since at low temperatures the thermal impedences are dominated by the contacts between dissimilar materials rather than by the conduction along the bulk, the order of magnitude of the thermal conductance is fixed by the geometry. These considerations show that the Cu block solution looks the most viable, providing in addition an easy method to tune the pulse modification effects just acting on the block volume.

\section{Conclusions and future prospects}
\label{sec:conclusion}

In conclusion, the ability to discriminate surface events from bulk events has been demonstrated experimentally and a thermal model describing the system has been developed. The model agrees with experimental measurements qualitatively and future application for bolometric experiments seems promising. Shield events can be distinguished from absorber events through amplitude comparison of the simultaneous pulses in the main absorber and the active shields. Reading out the shield thermistors in parallel is an effective way of reducing the total number of acquisition channels without compromising the event location discrimination of the SSBs, provided that reasonable reproducibility in the thermal couplings is achieved. 

Pulse shape discrimination can help discriminate main absorber from background events and reduce the overall number of acquisition channels further: two different ways could be explored. Event location can be identified through pulse shape discrimination of the active shields read out in parallel, in principle eliminating the needs for the main-absorber read-out. This method can differentiate between shield and mixed events from main absorber ones through the different resultant rise times, and would provide information on the energy deposited in the main absorber through the slow pulses; it probably reduces the overall energy resolution of main absorber events, but is worth being studied and optimized. The other way, acquiring only the signal coming from the main thermistor, seems promising as well by differentiating by the event decay time. The shields would act as pulse-shape modifiers, without the complication of additional read-out channels from independent thermistors. The addition of a thermal load on the shield can enhance this effect. This will be very useful in large-scale experiments for which surface-event discrimination is necessary as much as the lowering of the overall number of readout channels.

Active discrimination of surface background may have a large impact on the sensitivity of next-generation CUORE-like double beta decay experiments. For the CUORE experiment, the background in the $0\nu\beta\beta$ region has been extensively studied with Monte Carlo simulations, taking into account all anticipated sources (bulk and surface radioactive impurities, cosmic-ray muons, neutrons, etc.)~\cite{CUORE}. The results show that the present configuration with no SSB has a background level of about 10 counts/keV/ton/y, providing a sensitivity to the neutrinoless double-beta-decay half-life of $^{130}$Te of about 2$\times$10$^{26}$~y. This sensitivity would just reach the inverted hierarchy region of the neutrino mass pattern \cite{RODIN, SUHO, POVES}. By eliminating surface events, an additional factor of 10 reduction in background can be achieved according to simulations. In terms of sensitivity to Majorana neutrino mass, this would extend the search deep into the inverted hierarchy region and substantially increase the discovery potential of the experiment. 

\section{Acknowledgements}

We are pleased to acknowledge that this work has been partially supported by the ILIAS integrating activity (Contract No. RII3-CT-2004-506222) as part of the European Union FP6 programme in Astroparticle Physics, and by the PRIN-2006 project ``Optimization of Bolometric Detectors for Neutrino Physics'' (Italian Ministery of Research and University).

\bibliographystyle{elsarticle-num}

\end{document}